\documentclass[10pt]{article}
\usepackage[T1,T2A]{fontenc}
\usepackage[cp1251]{inputenc}
\usepackage{mathtools}
\usepackage{amsmath}
\usepackage{amssymb}
\usepackage{stackrel}
\usepackage{esint}


\begin{document}
\title{Classical first passage problems for $p$-adic stochastic processes}
\author{A.\,Kh.~Bikulov \\
 \textit{N.N. Semenov Federal Research Center for Chemical Physics} \\
  \textit{Russian Academy of Sciences,} \\
 \textit{Kosygin street 4, 117734 Moscow, Russia} \\
 e-mail:\:\texttt{beecul@mail.ru} \\
 and \\
 A.\,P.~Zubarev \\
 \textit{ Physics Department, Samara University, } \\
 \textit{ Moskovskoe shosse 34, 443123, Samara, Russia} \\
 \textit{Natural Science Department, } \\
 \textit{Samara State University of Railway Transport,} \\
 \textit{Perviy Bezimyaniy pereulok 18, 443066, Samara, Russia} \\
 e-mail:\:\texttt{apzubarev@mail.ru} }
\maketitle
\begin{abstract}
In this paper we present a comprehensive analysis of the solution
of the classical problem of finding the distribution density of a
random variable -- the first passage time to a given domain by the
trajectory of a $p$-adic Markov stochastic process with
probability density function satisfying the solution of the Cauchy
problem for the Vladimirov equation (a $p$-adic analog of the
Kolmogorov--Feller equation with the kernel of the Vladimirov
operator) with uniform initial distribution in the unit ball. We
consider three equivalent approaches to obtain equations for the
distribution density of a random variable  -- the first passage
time to a given domain by a stochastic trajectory. We find a
solution to these equations for the distribution density, analyze
its properties, and compare them with the properties of the
distribution density of a random variable -- the first return time
of a stochastic trajectory to the support of the initial
distribution. We also solve the problem of finding the number of
hittings a given domain and analyze the solution obtained. In
conclusion, we discuss a class of problems related to the study of
the distribution density of the passage time to a given domain and
the return time to the initial domain for other types of $p$-adic
Markov stochastic processes.
\end{abstract}

\section{Introduction}

$p$-Adic mathematical physics is an adequate mathematical
apparatus for describing various hierarchically organized complex
systems (see the surveys \cite{ALL,ALL_1,DKKM} and references
therein). One of the directions of research in this field, to
which many papers have been devoted, is the study of various
stochastic processes on $p$-adic and more general ultrametric
spaces. The present study also belongs to this field of research
and is a natural continuation of one of the previous studies
\cite{ABZ_2009} of the authors in which they solved the problem of
the distribution of the first return time to the support of the
initial distribution for a $p$-adic random walk whose distribution
density is described by the Vladimirov equation \cite{VVZ}. In the
subsequent paper \cite{B_2010_1} of the first named author, the
problem of the first passage to a given domain was analyzed for a
$p$-adic random walk trajectory. In this case the distribution
densities of the $p$-adic stochastic process satisfied the Cauchy
problem with the time-shifted initial condition. Moreover, in this
paper the author solved the problem of finding the average number
of hittings a given domain. The solution of this problem was
motivated by the need to describe experiments on spectral
diffusion in proteins. Here we present a comprehensive study of
the classical problem of finding the distribution density of the
first passage time of the trajectory of a $p$-adic random walk to
a ball of arbitrary radius that does not intersect the support of
the initial distribution.

In the present study, we apply the standard mathematical apparatus
to the analysis of complex-valued functions defined on the field
of $p$-adic numbers (see \cite{VVZ} for details). Here we give a
list of notations used in the paper. $\mathbb{Q}_{p}$ is a field
of $p$-adic numbers, and $\left|x\right|_{p}$ is a $p$-adic norm
of a $p$-adic number $x\in\mathbb{Q}_{p}$. The set
$B_{r}\left(a\right)=\left\{
x\in\mathbb{Q}_{p}\vcentcolon\left|x-a\right|_{p}\leq
p^{r}\right\} $ is a $p$-adic ball of radius $p^{r}$ centered at
point $a$. The set $S_{r}\left(a\right)$ $=\left\{
x\in\mathbb{Q}_{p}\vcentcolon\left|x-a\right|_{p}=p^{r}\right\} $
is a $p$-adic sphere of radius $p^{r}$ with center at $a$. Since
the field $\mathbb{Q}_{p}$ forms a locally compact additive
Abelian group, it follows that there exists a unique (up to a
factor) Haar measure in $\mathbb{Q}_{p}$ $d_{p}x$ that is
translation invariant: $d_{p}\left(x+a\right)=d_{p}x$. This
measure is usually normalized as

\[
\underset{\mathbb{Z}_{p}}{\int}d_{p}x=1,
\]
where $\mathbb{Z}_{p}\equiv B_{0}\left(0\right)$ is the ring of
$p$-adic integers. The Fourier transform of a function
$\varphi\left(x\right)\in L^{1}\left(\mathbb{Q}_{p},d_{p}x\right)$
is defined by

\[
\tilde{\varphi}\left(x\right)=\underset{\mathbb{Q}_{p}}{\int}\varphi\left(k\right)\chi\left(kx\right)d_{p}x\;x\in\mathbb{Q}_{p},
\]
where $\chi\left(x\right)$ is a normalized additive character of
the field $\mathbb{Q}_{p}$. In this case the inverse Fourier
transform is given by
\[
\varphi\left(x\right)=\underset{\mathbb{Q}_{p}}{\int}\tilde{\varphi}\left(k\right)\chi\left(-kx\right)d_{p}x\;x\in\mathbb{Q}_{p}
\]
if $\tilde{\varphi}\left(k\right)\in
L^{1}\left(\mathbb{Q}_{p},d_{p}k\right)$. The Vladimirov
pseudodifferential operator $D_{x}^{\alpha}$ has the form

\begin{equation}
D_{x}^{\alpha}\varphi\left(x\right)=\dfrac{1}{\Gamma_{p}\left(-\alpha\right)}\underset{\mathbb{Q}_{p}}{\int}\dfrac{\varphi\left(y\right)-\varphi\left(x\right)}{\left|x-y\right|_{p}^{\alpha+1}}d_{p}x,\label{D}
\end{equation}
where
$\Gamma_{p}\left(-\alpha\right)=\dfrac{1-p^{-\alpha-1}}{1-p^{\alpha}}$
is a $p$-adic analog of the gamma function. The operator (\ref{D})
for real numbers $\alpha>0$ is defined on complex-valued functions
of class $W^{a}$ with $0\leq a<\alpha$. This class of functions
$W^{a}$ $\left(a\geq0\right)$ is defined by the following
conditions:

(1) $\left|\varphi\left(x\right)\right|\leq
C\left(1+\left|x\right|_{p}^{a}\right)$, where $C$ is a real
number;

(2) there exists a positive integer $l=l\left(\varphi\right)$ such
that the following equality holds for any $x\in\mathbb{Q}_{p}$ and
any $x^{\prime}\in\mathbb{Q}_{p}$ such that
$\left|x^{\prime}\right|_{p}\leq p^{-l}$:
$\varphi\left(x+x^{\prime}\right)=\varphi\left(x\right)$.

The Vladimirov equation \cite{VVZ} has the form of a parabolic
equation with the right-hand side the pseudodifferential operator
(\ref{D}):
\begin{equation}
\dfrac{\partial}{\partial t}\varphi\left(x,t\right)=-D_{x}^{\alpha}\varphi\left(x,t\right).\label{p-KF}
\end{equation}

Let $\Omega$ be a set of elementary events, $\Sigma$ be the sigma
algebra of the subsets of $\Omega$, and $\mathrm{P}$ be a
countably additive nonnegative measure on $\Sigma$ satisfying the
condition $\mathrm{P}\left(\Omega\right)=1$. Next, let
$\mathfrak{B}$ be the sigma algebra of the subsets of
$\mathbb{Q}_{p}$. A map
$\xi\vcentcolon\Omega\rightarrow\mathbb{Q}_{p}$ is
$\Sigma\mid\mathfrak{B}$-measurable if
$\xi^{-1}\left(\mathfrak{B}\right)\subset\Sigma.$ In this case the
$\Sigma\mid\mathfrak{B}$-measurable map is called a $p$-adic
random variable $\xi=\xi\left(\omega\right)$. The function
$\xi\left(\omega\right)$ induces a probability measure
$P_{\xi}\left(B\right)=\mathrm{P}\left\{
\xi^{-1}\left(B\right)\right\} $ on the set $B\in\mathfrak{B}$,
and $P_{\xi}\left(B\right)$ is the distribution function of the
random variable $\xi\left(\omega\right)$. A stochastic process on
$\mathbb{Q}_{p}$ is a map
$\xi\left(t,\omega\right)\colon\:\mathbb{R}_{+}\times\Omega\rightarrow\mathbb{Q}_{p}$
that, for an arbitrary fixed $t\in\mathbb{R}_{+}$, is a measurable
map from $\left(\Omega,\Sigma\right)$ into
$\left(\mathbb{Q}_{p},\mathfrak{B}\right)$, i.e., a map such that

\[
\left\{ \omega\in\Omega\vcentcolon\xi\left(t,\omega\right)\in
B\right\} \in\Sigma
\]
for any $B\in\mathfrak{B}$.

In this study we present a comprehensive analysis of the solution
to the classical problem of finding the distribution density of
the first passage time to a ball $B_{r}\left(a\right)$ of radius
$p^{r}$ with center at an arbitrary point
$a\in\mathbb{Q}_{p}\setminus\mathbb{Z}_{p}$ by a $p$-adic Markov
stochastic process $\xi\left(t,\omega\right)$ whose distribution
function is the solution of the Cauchy problem for the Vladimirov
equation (\ref{p-KF}) with the initial condition given by a
uniform distribution in $\mathbb{Z}_{p}$:

\begin{equation}
\varphi\left(x,0\right)=\Omega\left(\left|x\right|_{p}\right),\;\Omega\left(\lambda\right)=\begin{cases}
1, & \lambda\leq1,\\
0 & \lambda>1.
\end{cases}\label{I_C}
\end{equation}
Here we consider three equivalent approaches to obtain an equation
for the distribution density function
$f\left(t\right)\colon\mathbb{R}_{+}\rightarrow\mathbb{R}_{+}$ of
random variable $\tau_{B_{r}\left(a\right)}\left(\omega\right)$ --
the first passage time of the homogeneous Markov process
$\xi\left(t,\omega\right)\colon\Omega\times\mathbb{R}_{+}\rightarrow\mathbb{Q}_{p}$
to the domain $B_{r}\left(a\right)\in\mathbb{Q}_{p}$. Further we
analyze the properties of the function $f\left(t\right)$ and
compare them with the properties of the probability density
function $f_{ret}\left(t\right)$ of a random variable
$\tau_{B_{r}}\left(\omega\right)$ -- the first return time to the
domain $B_{r}$, which is the support of the initial distribution.
In addition, we solve the problem of the distribution of the
number of passages to the domain $B_{r}\left(a\right)$.

The paper is organized as follows. In the next section we obtain
various forms of equation for the distribution density
$f\left(t\right)$ of the first passage time to the domain
$B_{r}\left(a\right)$ by the trajectory of the stochastic process
$\xi\left(t,\omega\right)$ and find its solution. In Section 3 we
analyze the properties of the function $f\left(t\right)$ and find
its asymptotic behavior as $t\rightarrow\infty$. In Section 4 we
solve the problem of the number of hittings the domain
$B_{r}\left(a\right)$ by the trajectories of the stochastic
process $\xi\left(t,\omega\right)$  The concluding section is
devoted to the discussion of the class of problems related to the
study of the distribution density of the first passage time to a
given domain for other types of $p$-adic Markov stochastic
processes with the initial condition on a compact set.

\section{Equations for the distribution density of the first passage time}

We consider a $p$-adic-valued stochastic process
$\xi\left(t,\omega\right)\vcentcolon\mathbb{R}_{+}\times\Omega\rightarrow\mathbb{Q}_{p}$
whose distribution density is the solution of the Cauchy problem
for the Vladimirov equation (\ref{p-KF}) with the initial
condition (\ref{I_C}). This stochastic process is a homogeneous
Markov process continuous by the Kolmogorov condition \cite{VVZ}.

In \cite{ABZ_2009}, the authors posed and solved the problem of
finding the distribution density of a random variable
$\tau_{\mathbb{Z}_{P}}\left(\omega\right)$ -- the first return
time of the trajectory of a stochastic process
$\xi\left(t,\omega\right)$ to the domain of the initial condition
$\mathbb{Z}_{p}$. In this paper the authors proposed two
equivalent approaches to finding an equation for the distribution
density function $f_{ret}\left(t\right)$ of the random variable
$\tau_{\mathbb{Z}_{P}}\left(\omega\right)$. One of these
approaches is based on probability analysis and immediately leads
to an equation for $f_{ret}\left(t\right)$, which is an
inhomogeneous Volterra integral equation of the second kind. The
other approach leads to the equation for $f_{ret}\left(t\right)$,
which is expressed in terms of the solution of the Cauchy problem
for the pseudodifferential equation for a $p$-adic random walk
with an absorbing region. In this section  we also apply the above
approaches to find the corresponding equations for the
distribution density $f\left(t\right)$ of the random variable
$\tau_{B_{r}\left(a\right)}\left(\omega\right)$ -- the first
passage time to the domain $B_{r}\left(a\right)\in\mathbb{Q}_{p}$,
$B_{r}\left(a\right)\cap\mathbb{Z}_{p}=\textrm{\O}$.

\textbf{Definition.} The first passage time of the trajectory of a
stochastic process $\xi\left(t,\omega\right)$ to the domain
$B_{r}\left(a\right)\in\mathbb{Q}_{p}$,
$B_{r}\left(a\right)\cap\mathbb{Z}_{p}=\textrm{\O}$, is a random
variable
$\tau_{B_{r}\left(a\right)}\left(\omega\right)\vcentcolon\Omega\rightarrow\mathbb{R}_{+}$
defined by:

\begin{equation}
\tau_{B_{r}\left(a\right)}\left(\omega\right)=\inf\left\{ t>0\vcentcolon\xi\left(t,\omega\right)\in B_{r}\left(a\right),\mathrm{if}\:\xi\left(0,\omega\right)\in\mathbb{Z}_{p}\right\} .\label{def rv}
\end{equation}

The initial condition for the Cauchy problem (\ref{I_C}) yields

\[
\mathrm{P}\left\{
\omega\in\Omega\vcentcolon\left|\xi\left(0,\omega\right)\right|_{p}\leq1\right\}
=1.
\]

\textbf{Theorem 1.} The distribution density $f\left(t\right)$ of
the random variable
$\tau_{B_{r}\left(a\right)}\left(\omega\right)$ under the
condition $\left|a\right|_{p}>p^{r}$ satisfies the Volterra
equation of the second kind

\begin{equation}
g\left(t\right)=\stackrel[0]{t}{\int}g_{R}\left(t-\tau\right)f\left(\tau\right)d\tau+f\left(t\right),\label{Volt_2}
\end{equation}
where
\begin{equation}
g\left(t\right)=-\dfrac{1}{\Gamma_{p}\left(-\alpha\right)}\underset{B_{r}\left(a\right)}{\int}d_{p}y\underset{\mathbb{Q}_{p}\setminus B_{r}\left(a\right)}{\int}\dfrac{\varphi\left(x,t\right)}{\left|x-y\right|_{p}^{\alpha+1}}d_{p}x,\label{g(t)_T1}
\end{equation}
 $\varphi\left(x,t\right)$ is a solution of the Cauchy problem   (\ref{p-KF})
--(\ref{I_C}), and the function $g_{R}\left(t\right)$ is given by
\begin{equation}
g_{R}\left(t\right)=-\dfrac{1}{\Gamma_{p}\left(-\alpha\right)}\underset{B_{r}\left(a\right)}{\int}d_{p}y\underset{\mathbb{Q}_{p}\setminus B_{r}\left(a\right)}{\int}\dfrac{\psi\left(x,t\right)}{\left|x-y\right|_{p}^{\alpha+1}}d_{p}x,\label{g(t)_R_T1}
\end{equation}
where $\psi\left(x,t\right)$ is a solution of the Cauchy problem
for equation (\ref{p-KF}) with the initial condition

\begin{equation}
\psi\left(x,0\right)=p^{-r}\Omega\left(\left|x-a\right|_{p}p^{-r}\right).\label{NU_B}
\end{equation}

\textbf{Proof. } Let $A\left(t_{i}\right)$ be an event such that
the trajectory of the process $\xi\left(t,\omega\right)$ reaches
the domain $B_{r}\left(a\right)$ at time $t_{i}$:

\[
A\left(t_{i}\right)=\left\{ \omega\in\Omega\vcentcolon\xi\left(t_{i},\omega\right)\in B_{r}\left(a\right),if\:\exists\:t<t_{i}\vcentcolon\forall\:\tau\in\left(t,t_{i}\right)\:\xi\left(\tau,\omega\right)\in\mathbb{Q}_{p}\setminus B_{r}\left(a\right)\right\} .
\]
Let $B\left(t_{i-1},t_{i}\right)$ be an event such that the
trajectory of the process $\xi\left(t,\omega\right)$ reaches the
domain $B_{r}\left(a\right)$ at time
$t\in\left(t_{i-1},t_{i}\right]$ for the first time:

\[
B\left(t_{i-1},t_{i}\right)=\left\{ \omega\in\Omega\vcentcolon t_{i-1}<\tau_{B_{r}\left(a\right)}\left(\omega\right)\leq t_{i}\right\} .
\]
Let us divide the segment $\left[0,t\right]$ into $n$ parts by the
points $0\equiv t_{0}<t_{1}<t_{2}<\ldots<t_{n-1}<t_{n}\equiv t$.
It is obvious that
$A\left(t_{n}\right)\subset\bigcup_{i=1}^{n-1}B\left(t_{i-1},t_{i}\right)$
and, since $A\left(t_{n}\right)\cap
B\left(t_{n-1},t_{n}\right)=B\left(t_{n-1},t_{n}\right)$, it
follows that

\begin{equation}
A\left(t_{n}\right)=A\left(t_{n}\right)\cap\left(\bigcup_{i=1}^{n-1}B\left(t_{i-1},t_{i}\right)\right)=\bigcup_{i=1}^{n-1}\left(A\left(t_{n}\right)\cap B\left(t_{i-1},t_{i}\right)\right)\cup B\left(t_{n-1},t_{n}\right).\label{A(t_n)}
\end{equation}
Let $\mathrm{P}\left\{ A\left(t_{n}\right)\right\} $ and
$\mathrm{P}\left\{ B\left(t_{i-1},t_{i}\right)\right\} $ be the
probabilities of the events $A\left(t_{n}\right)$ and
$B\left(t_{i-1},t_{i}\right)$. Then, taking into account
(\ref{A(t_n)}) and the fact that the events
$B\left(t_{i-1},t_{i}\right)$ are incompatible, we write

\[
\mathrm{P}\left\{ A\left(t_{n}\right)\right\}
=\sum_{i=1}^{n-1}\mathrm{P}\left\{ A\left(t_{n}\right)\cap
B\left(t_{i-1},t_{i}\right)\right\} +\mathrm{P}\left\{
B\left(t_{n-1},t_{n}\right)\right\},
\]

\begin{equation}
\mathrm{P}\left\{ A\left(t_{n}\right)\right\}
=\sum_{i=1}^{n-1}\mathrm{P}\left\{ A\left(t_{n}\right)\mid
B\left(t_{i-1},t_{i}\right)\right\} \mathrm{P}\left\{
B\left(t_{i-1},t_{i}\right)\right\} +\mathrm{P}\left\{
B\left(t_{n-1},t_{n}\right)\right\}. \label{P event}
\end{equation}
Since the process $\xi\left(t,\omega\right)$ is homogeneous and
Markovian, we have
\[
\mathrm{P}\left\{ A\left(t\right)|B\left(t_{i-1},t_{i}\right)\right\} =\mathrm{P}\left\{ A\left(t-t_{i}\right)|B\left(0,t_{i}-t_{i-1}\right)\right\}
\]

\begin{equation}
=-\dfrac{1}{\Gamma_{p}}\underset{B_{r}\left(a\right)}{\int}d_{p}y\underset{\mathbb{Q}_{p}\setminus B_{r}\left(a\right)}{\int}\dfrac{\psi\left(x,t-t_{i}\right)}{\left|x-y\right|_{p}^{\alpha+1}}d_{p}x+O\left(t_{i}-t_{i-1}\right)=g_{R}\left(t-t_{i}\right)+O\left(t_{i}-t_{i-1}\right),\label{P(A|B)}
\end{equation}
where we used the conventional notation $O\left(\Delta t\right)$
for the infinitesimal function as $\Delta t\rightarrow0$. It is
obvious that

\begin{equation}
\mathrm{P}\left\{ A\left(t\right)\right\}
=-\dfrac{1}{\Gamma_{p}}\underset{B_{r}\left(a\right)}{\int}d_{p}y\underset{\mathbb{Q}_{p}\setminus
B_{r}\left(a\right)}{\int}\dfrac{\varphi\left(x,t\right)}{\left|x-y\right|_{p}^{\alpha+1}}d_{p}x=g\left(t\right);\label{P(A)}
\end{equation}
moreover,

\begin{equation}
\mathrm{P}\left\{ B\left(t_{i},t_{i-1}\right)\right\} =\mathrm{P}\left\{ B\left(0,t_{i}-t_{i-1}\right)\right\} =f\left(t_{i}\right)\left(t_{i}-t_{i-1}\right)+o\left(t_{i}-t_{i-1}\right),\label{P(B)}
\end{equation}
where $o\left(\Delta t\right)$ denotes a function such that
$\lim_{\Delta t\rightarrow0}\dfrac{o\left(\Delta t\right)}{\Delta
t}=0$. Then, taking into account (\ref{P(A|B)}), (\ref{P(A)}), and
(\ref{P(B)}), we obtain the following equation from (\ref{P
event}):
\begin{equation}
g\left(t\right)=\sum_{i=1}^{n-1}\left(g_{R}\left(t-t_{i}\right)+O\left(t_{i}-t_{i-1}\right)\right)\left(f\left(t_{i}\right)\left(t_{i}-t_{i-1}\right)+o\left(t_{i}-t_{i-1}\right)\right)+f\left(t\right)\left(t-t_{n-1}\right)+o\left(t-t_{n-1}\right).\label{g(t)}
\end{equation}
As $n\rightarrow\infty$ and $\underset{i}{\max}\left\{
\left|t_{i}-t_{i-1}\right|\right\} \rightarrow0$, from
(\ref{g(t)}) we obtain (\ref{Volt_2}), which completes the proof
of Theorem 1.

Let us show that the distribution density $f\left(t\right)$
satisfies the Volterra equation of the first kind.

\textbf{Theorem 2.} The distribution density $f\left(t\right)$ of
the random variable
$\tau_{B_{r}\left(a\right)}\left(\omega\right)$ satisfies the
Volterra integral equation of the first kind

\begin{equation}
S_{r}\left(t\right)=\intop_{0}^{t}S\left(t-t^{\prime}\right)f\left(t^{\prime}\right)dt^{\prime},\label{Volt_1}
\end{equation}
where
\begin{equation}
S\left(t\right)=\intop_{B_{r}\left(a\right)}d_{p}x\psi\left(x,t\right),\label{SP}
\end{equation}

\begin{equation}
S_{r}\left(t\right)=\intop_{B_{r}\left(a\right)}d_{p}x\varphi\left(x,t\right),\label{SPR}
\end{equation}
and $\varphi\left(x,t\right)$ and $\psi\left(x,t\right)$ are the
solutions of the Cauchy problem for equation (\ref{p-KF}) with the
initial conditions
$\varphi\left(x,0\right)=\Omega\left(\left|x\right|_{p}\right)$
and
$\psi\left(x,0\right)=\dfrac{1}{p^{r}}\Omega\left(\left|x-a\right|_{p}p^{-r}\right)$,
respectively.

\textbf{Proof. } Suppose that $A\left(t\right)$ is an event such
that the trajectory of the stochastic process
$\xi\left(t,\omega\right)$ is in the domain $B_{r}\left(a\right)$
at time $t$:

\[
A\left(t\right)=\left\{ \omega\in\Omega\vcentcolon\xi\left(t,\omega\right)\in B_{r}\left(a\right)\right\} .
\]
Let us split the time interval $\left(0,t\right]$ into $n$
intervals $\left(t_{0},t_{1}\right]$, $\left(t_{1},t_{2}\right]$,
$\ldots$, $\left(t_{n-1},t_{n}\right]$, where $t_{0}=0$ and
$t_{n}=t$, and denote $\Delta t_{i}=t_{i}-t_{i-1}$. Suppose that
the event $B\left(t_{i-1},t_{i}\right)$ ($i=1,2,\ldots,n$) is such
that the trajectory of the stochastic process
$\xi\left(t,\omega\right)$ reaches the domain
$B_{r}\left(a\right)$ for the first time in the interval
$\left(t_{i-1},t_{i}\right]$:

\[
B\left(t_{i-1},t_{i}\right)=\left\{ \omega\in\Omega\vcentcolon t_{i-1}<\tau_{B_{r}\left(a\right)}\left(\omega\right)\leq t_{i}\right\} .
\]
Then
\[
A\left(t\right)=A\left(t\right)\cap\left(\bigcup_{i=1}^{n}B\left(t_{i-1},t_{i}\right)\right)=\bigcup_{i=1}^{n}\left(A\left(t\right)\cap B\left(t_{i-1},t_{i}\right)\right)
\]
and
\[
\mathrm{P}\left\{ A\left(t\right)\right\} =\mathrm{P}\left\{ \bigcup_{i=1}^{n}\left(A\left(t\right)\cap B\left(t_{i-1},t_{i}\right)\right)\right\}
\]
\begin{equation}
=\sum_{i=1}^{n}\mathrm{P}\left\{ A\left(t\right)\cap B\left(t_{i-1},t_{i}\right)\right\} =\sum_{i=1}^{n}\mathrm{P}\left\{ A\left(t\right)|B\left(t_{i-1},t_{i}\right)\right\} \mathrm{P}\left\{ B\left(t_{i-1},t_{i}\right)\right\} .\label{P}
\end{equation}
Let $\psi\left(x,t\right)$ be the solution of the Cauchy problem
for equation (\ref{p-KF}) with the initial condition
$\psi\left(x,0\right)=\dfrac{1}{p^{r}}\Omega\left(\left|x-a\right|_{p}p^{-r}\right)$.
Then
\begin{equation}
\mathrm{P}\left\{ A\left(t\right)\right\} =\underset{B_{r}\left(a\right)}{\int}\varphi\left(x,t\right)d_{p}x=S_{r}\left(t\right).\label{z1}
\end{equation}
Next, since the process $\xi\left(t,\omega\right)$ is homogeneous
and Markovian, we have

\[
\mathrm{P}\left\{ A\left(t\right)|B\left(t_{i-1},t_{i}\right)\right\} =\mathrm{P}\left\{ A\left(t-t_{i}\right)|B\left(0,t_{i}-t_{i-1}\right)\right\}
\]

\begin{equation}
=\underset{B_{r}\left(a\right)}{\int}\psi\left(x,t-t_{i}\right)d_{p}x+O\left(t_{i}-t_{i-1}\right)=S\left(t-t_{i}\right)+O\left(t_{i}-t_{i-1}\right).\label{z2}
\end{equation}
Moreover, we have
\begin{equation}
\mathrm{P}\left\{ B\left(t_{i-1},t_{i}\right)\right\} =f\left(t_{i}\right)\left(t_{i}-t_{i-1}\right)+o\left(t_{i}-t_{i-1}\right).\label{z3}
\end{equation}
Thus, equation (\ref{P}) with regard to (\ref{z1}) -- (\ref{z3})
implies

\begin{equation}
S_{r}\left(t\right)=\sum_{i=1}^{n}\left(S\left(t-t_{i}\right)+O\left(t_{i}-t_{i-1}\right)\right)\left(f\left(t_{i}\right)\left(t_{i}-t_{i-1}\right)+o\left(t_{i}-t_{i-1}\right)\right)\label{PP}
\end{equation}
In the limit as $n\rightarrow\infty$ and $\max_{i}\left\{
\left|t_{i}-t_{i-1}\right|\right\} \rightarrow0$, (\ref{PP}) turns
into (\ref{Volt_1}), which proves Theorem 2.

Note that, since the functions $S\left(t\right)$ and
$\dfrac{d}{dt}S\left(t\right)$ are continuous, the Volterra
integral equation of the first kind (\ref{Volt_1}) is reduced by
differentiation to the Volterra integral equation of the second
kind (\ref{Volt_2}). Indeed, after differentiating (\ref{Volt_1})
we have

\begin{equation}
\dfrac{d}{dt}S_{r}\left(t\right)=S\left(0\right)f\left(t\right)+\stackrel[0]{t}{\int}\dfrac{d}{dt}S\left(t-\tau\right)f\left(\tau\right)d\tau.\label{V-a1}
\end{equation}
Integrating both sides of equation (\ref{p-KF}) with the solutions
$\varphi\left(x,t\right)$ and $\psi\left(x,t\right)$ over the
domain $B_{r}\left(a\right)$, we write

\begin{equation}
\dfrac{d}{dt}S_{r}\left(t\right)=-B_{\alpha}\left(r\right)S_{r}\left(t\right)+g\left(t\right)\label{vl-1}
\end{equation}
and

\begin{equation}
\dfrac{d}{dt}S\left(t\right)=-B_{\alpha}\left(r\right)S\left(t\right)+g_{R}\left(t\right),\label{vl-2}
\end{equation}
where

\begin{equation}
B_{\alpha}\left(r\right)=p^{-\alpha r}\dfrac{1-p^{-1}}{1-p^{-\alpha-1}}.\label{B_alpha_r}
\end{equation}
Substituting (\ref{vl-1}) and (\ref{vl-2}) into (\ref{V-a1}) and
taking into account that $S\left(0\right)=1$, we obtain
(\ref{Volt_2}).

Below when analyzing the properties of the distribution density
$f\left(t\right)$ of the random variable
$\tau_{B_{r}\left(a\right)}\left(\omega\right)$ -- the first
passage time to the domain $B_{r}\left(a\right)$ -- we will
compare them with the properties of the distribution density
$f_{ret}\left(t\right)$ of the random variable
$\tau_{\mathbb{Z}_{p}}\left(\omega\right)$ -- the first return
time to the domain $\mathbb{Z}_{p}$. In  \cite{ABZ_2009}, the
authors applied the approach based on probability analysis and
obtained an equation for the probability density of the random
variable $\tau_{\mathbb{Z}_{p}}\left(\omega\right)$ in the form of
the Volterra integral equation of the second kind:

\begin{equation}
g\left(t\right)=\stackrel[0]{t}{\int}g\left(t-\tau\right)f_{ret}\left(\tau\right)d\tau+f_{ret}\left(t\right),\label{Volt_2_ret}
\end{equation}
where
$g\left(t\right)=-\dfrac{1}{\Gamma_{p}}\underset{\mathbb{Z}_{p}}{\int}d_{p}y\underset{\mathbb{Q}_{p}\setminus\mathbb{Z}_{p}}{\int}\dfrac{\varphi\left(x,t\right)}{\left|x-y\right|_{p}^{\alpha+1}}d_{p}x$,
and $\varphi\left(x,t\right)$ is the solution of the Cauchy
problem (\ref{p-KF}) -- (\ref{I_C}). Below in Theorem 3 we apply
probability arguments to show that the equation for the function
$f_{ret}\left(t\right)$ can also be expressed in the form of the
Volterra equation of the first kind, which is equivalent to
equation (\ref{Volt_2_ret}).

\textbf{Theorem 3.} Let $\varphi\left(x,t\right)$ be the solution
of the Cauchy problem (\ref{p-KF}) -- (\ref{I_C}). Then the
distribution density function $f_{ret}\left(t\right)$ satisfies
the Volterra equation of the first kind

\begin{equation}
S\left(t\right)=\exp\left(-B_{\alpha}t\right)+\intop_{0}^{t}S\left(t-\tau\right)f_{ret}\left(\tau\right)d\tau,\label{Volt_1_ret}
\end{equation}
where
\[
S_{\mathbb{Z}_{p}}\left(t\right)=\intop_{\mathbb{Z}_{p}}\varphi\left(x,t\right)d_{p}x
\]
and
\[
B_{\alpha}=-\dfrac{1}{\Gamma_{p}\left(-\alpha\right)}\intop_{\mathbb{Q}_{p}\setminus\mathbb{Z}_{p}}d_{p}x\intop_{\mathbb{Z}_{p}}d_{p}y\dfrac{1}{\left|x-y\right|_{p}^{\alpha+1}}=\dfrac{1-p^{-1}}{1-p^{-\alpha-1}}.
\]

\textbf{Proof. } Suppose that the event $C\left(t\right)$ is such
that the trajectory in the interval $\left(0,t\right]$ does not
leave $\mathbb{Z}_{p}$:

\[
C\left(t\right)=\left\{ \omega\in\Omega\vcentcolon\xi\left(\tau,\omega\right)\in\mathbb{Z}_{p}\forall0<\tau\leq t\right\} .
\]
Then the probability of this event is
\[
\mathrm{P}\left\{ C\left(t\right)\right\} =\exp\left(-B_{\alpha}t\right).
\]
Let us divide the time interval $\left(0,t\right]$ into $n$
intervals $\left(t_{0},t_{1}\right]$, $\left(t_{1},t_{2}\right]$,
$\ldots$, and $\left(t_{n-1},t_{n}\right]$, where $t_{0}=0$ and
$t_{n}=t$, and denote $\Delta t_{i}=t_{i}-t_{i-1}$. Suppose that
the event $A\left(t\right)$ ($i=1,2,\ldots,n$) is such that the
trajectory at time $t$ is in the domain $\mathbb{Z}_{p}$,

\[
A\left(t\right)=\left\{ \omega\in\Omega\vcentcolon\xi\left(t,\omega\right)\in\mathbb{Z}_{p}\right\} .
\]
Suppose that the event $B\left(t_{i-1},t_{i}\right)$
($i=1,2,\ldots,n$) is such that the trajectory returns to the
domain $\mathbb{Z}_{p}$ for the first time in the interval
$\left(t_{i-1},t_{i}\right]$,

\[
B\left(t_{i-1},t_{i}\right)=\left\{ \omega\in\Omega\vcentcolon t_{i-1}<\tau_{\mathbb{Z}_{p}}\left(\omega\right)\leq t_{i}\right\} .
\]
Then
\[
A\left(t\right)=A\left(t\right)\cap C\left(t\right)+A\left(t\right)\cap\overline{C}\left(t\right)
\]

\[
=C\left(t\right)+A\left(t\right)\cap\overline{C}\left(t\right)\cap\left(\bigcup_{i=1}^{n}B\left(t_{i-1},t_{i}\right)\right)
\]
\[
=C\left(t\right)+\bigcup_{i=1}^{n}\left(A\left(t\right)\cap\overline{C}\left(t\right)\cap B\left(t_{i-1},t_{i}\right)\right)
\]
\[
=C\left(t\right)+\bigcup_{i=1}^{n}\left(A\left(t\right)\cap B\left(t_{i-1},t_{i}\right)\right)
\]
and
\[
\mathrm{P}\left\{ A\left(t\right)\right\} =\mathrm{P}\left\{ C\left(t\right)\right\} +\mathrm{P}\left\{ \bigcup_{i=1}^{n}\left(A\left(t\right)\cap B\left(t_{i-1},t_{i}\right)\right)\right\}
\]

\[
=\mathrm{P}\left\{ C\left(t\right)\right\} +\sum_{i=1}^{n}\mathrm{P}\left\{ A\left(t\right)\cap B\left(t_{i-1},t_{i}\right)\right\}
\]

\begin{equation}
=\mathrm{P}\left\{ C\left(t\right)\right\} +\sum_{i=1}^{n}\mathrm{P}\left\{ A\left(t\right)|B\left(t_{i-1},t_{i}\right)\right\} \mathrm{P}\left\{ B\left(t_{i-1},t_{i}\right)\right\} .\label{P_ret}
\end{equation}
It is obvious that

\begin{equation}
\mathrm{P}\left\{ A\left(t\right)\right\} =\intop_{\mathbb{Z}_{p}}\varphi\left(x,t\right)d_{p}x=S_{\mathbb{Z}_{p}}\left(t\right).\label{zz1}
\end{equation}
Since the process $\xi\left(t,\omega\right)$ is homogeneous and
Markovian, we have

\[
\mathrm{P}\left\{ A\left(t\right)|B\left(t_{i-1},t_{i}\right)\right\} =\mathrm{P}\left\{ A\left(t-t_{i}\right)|B\left(0,t_{i}-t_{i-1}\right)\right\}
\]

\begin{equation}
=\intop_{\mathbb{Z}_{p}}\varphi\left(x,t-t_{i}\right)d_{p}x+O\left(t_{i}-t_{i-1}\right)=S_{\mathbb{Z}_{p}}\left(t-t_{i}\right)+O\left(t_{i}-t_{i-1}\right).\label{zz2}
\end{equation}
We also write
\begin{equation}
\mathrm{P}\left\{ B\left(t_{i},t_{i-1}\right)\right\} =f_{ret}\left(t_{i}\right)\left(t_{i}-t_{i-1}\right)+o\left(t_{i}-t_{i-1}\right).\label{zz3}
\end{equation}
With regard to (\ref{zz1}) -- (\ref{zz3}), (\ref{P_ret}) implies

\begin{equation}
S_{\mathbb{Z}_{p}}\left(t\right)=\exp\left(-B_{\alpha}t\right)+\sum_{i=1}^{n}\left(S_{\mathbb{Z}_{p}}\left(t-t_{i}\right)+O\left(t_{i}-t_{i-1}\right)\right)\left(f_{ret}\left(t_{i}\right)\left(t_{i}-t_{i-1}\right)+o\left(t_{i}-t_{i-1}\right)\right)\label{PPP}
\end{equation}
In the limit as $n\rightarrow\infty$ and $\max_{i}\left\{
\left|t_{i}-t_{i-1}\right|\right\} \rightarrow0$, (\ref{PPP})
turns into (\ref{Volt_1_ret}), which proves Theorem 3.

Note that, in exactly the same way as in the problem of the
distribution of the first passage time, the Volterra integral
equation of the first kind (\ref{Volt_1_ret}) is reduced by
differentiation to the Volterra integral equation of the second
kind (\ref{Volt_2_ret}).

It is known that in the classical problem of the random walk of a
particle on the real axis the distribution density function of the
first passage time to some domain can be found by solving the
diffusion equation with an absorbing region (see, for example,
\cite{Feller}). Let us show that this approach can also be applied
to finding the distribution density of
$\tau_{B_{r}\left(a\right)}\left(\omega\right)$.

\textbf{Theorem 4.} The distribution density $f\left(t\right)$ of
the random variable
$\tau_{B_{r}\left(a\right)}\left(\omega\right)$ has the form
\begin{equation}
f\left(t\right)=-\dfrac{1}{\Gamma_{p}\left(-\alpha\right)}\underset{B_{r}\left(a\right)}{\int}d_{p}x\underset{\mathbb{Q}_{p}\setminus B_{r}\left(a\right)}{\int}\dfrac{\phi\left(y,t\right)}{\left|y-x\right|_{p}^{\alpha+1}}d_{p}y,\label{sol_ret}
\end{equation}
where $\phi\left(x,t\right)$ is the solution of the Cauchy problem
for the $p$-adic random walk equation with an absorbing region in
$B_{r}\left(a\right)$,

\begin{equation}
\dfrac{\partial}{\partial
t}\phi\left(t,x\right)=-\dfrac{1}{\Gamma_{p}\left(-\alpha\right)}\left\{
\underset{\mathbb{Q}_{p}}{\int}\dfrac{\phi\left(y,t\right)-\phi\left(x,t\right)}{\left|x-y\right|_{p}^{\alpha+1}}d_{p}y-\Omega\left(\left|x-a\right|_{p}p^{-r}\right)\underset{\mathbb{Q}_{p}\setminus
B_{r}\left(a\right)}{\int}\dfrac{\phi\left(y,t\right)}{\left|x-y\right|_{p}^{\alpha+1}}d_{p}y\right\},
\label{Ab_ar}
\end{equation}
and the initial condition
$\phi\left(0,x\right)=\Omega\left(\left|x\right|\right)$.

\textbf{Proof. } Equation (\ref{Ab_ar}) can easily be obtained by
the following arguments. Let us write (\ref{p-KF}) for the
function $\phi\left(x,t\right)$,

\begin{equation}
\dfrac{d\phi\left(x,t\right)}{dt}=-\dfrac{1}{\Gamma_{p}\left(-\alpha\right)}\intop_{\mathbb{Q}_{p}}d_{p}y\dfrac{\phi\left(y,t\right)-\phi\left(x,t\right)}{\left|x-y\right|_{p}^{\alpha+1}},\label{eq_Vl_phi}
\end{equation}
and represent the function $\phi\left(x,t\right)$ as

\begin{equation}
\phi\left(x,t\right)=\phi_{1}\left(x,t\right)+\phi_{2}\left(x,t\right)\label{phi_1_2}
\end{equation}
where
$\phi_{1}\left(x,t\right)=\Omega\left(\left|x-a\right|_{p}p^{-r}\right)\phi\left(x,t\right)$
and
$\phi_{2}\left(x,t\right)=\left(1-\Omega\left(\left|x-a\right|_{p}p^{-r}\right)\right)\phi\left(x,t\right)$.
Substituting (\ref{phi_1_2}) into (\ref{eq_Vl_phi}) and projecting
equation (\ref{eq_Vl_phi}) onto the domains $B_{r}\left(a\right)$
and $\mathbb{Q}_{p}\setminus B_{r}\left(a\right)$, we obtain two
equations:

\[
\dfrac{\partial}{\partial t}\phi_{1}\left(x,t\right)=-\dfrac{1}{\Gamma_{p}\left(-\alpha\right)}\left[\underset{\mathbb{Q}_{p}\setminus B_{r}\left(a\right)}{\int}\dfrac{\phi_{2}\left(y,t\right)}{\left|y-x\right|_{p}^{\alpha+1}}d_{p}y\right.
\]

\[
\left.-\underset{\mathbb{Q}_{p}\setminus B_{r}\left(a\right)}{\int}\dfrac{\phi_{1}\left(x,t\right)}{\left|y-x\right|_{p}^{\alpha+1}}d_{p}y+\underset{B_{r}\left(a\right)}{\int}\dfrac{\phi_{1}\left(y,t\right)-\phi_{1}\left(x,t\right)}{\left|y-x\right|_{p}^{\alpha+1}}d_{p}y\right]
\]
for $x\in B_{r}\left(a\right)$ and

\[
\dfrac{\partial}{\partial t}\phi_{2}\left(x,t\right)=-\dfrac{1}{\Gamma_{p}\left(-\alpha\right)}\left[\underset{B_{r}\left(a\right)}{\int}\dfrac{\phi_{1}\left(y,t\right)}{\left|y-x\right|_{p}^{\alpha+1}}d_{p}y\right.
\]

\[
\left.-\underset{B_{r}\left(a\right)}{\int}\dfrac{\phi_{2}\left(x,t\right)}{\left|y-x\right|_{p}^{\alpha+1}}d_{p}y+\underset{\mathbb{Q}_{p}\setminus B_{r}\left(a\right)}{\int}\dfrac{\phi_{2}\left(y,t\right)-\phi_{2}\left(x,t\right)}{\left|y-x\right|_{p}^{\alpha+1}}d_{p}y\right]
\]
for $x\in\mathbb{Q}_{p}\setminus B_{r}\left(a\right)$. The first
term on the right-hand side of the first equation is the
probability of transition from the domain $\mathbb{Q}_{p}\setminus
B_{r}\left(a\right)$ to a point $x\in B_{r}\left(a\right)$ in unit
time. If we compensate for this term, namely, if we subtract

\[
-\dfrac{\Omega\left(\left|x-a\right|_{p}p^{-r}\right)}{\Gamma_{p}\left(-\alpha\right)}\underset{\mathbb{Q}_{p}\setminus B_{r}\left(a\right)}{\int}\dfrac{\phi_{2}\left(y,t\right)}{\left|y-x\right|_{p}^{\alpha+1}}d_{p}y
\]

\[
=-\dfrac{\Omega\left(\left|x-a\right|_{p}p^{-r}\right)}{\Gamma_{p}\left(-\alpha\right)}\underset{\mathbb{Q}_{p}\setminus B_{r}\left(a\right)}{\int}\dfrac{\phi_{1}\left(y,t\right)+\phi_{2}\left(y,t\right)}{\left|y-x\right|_{p}^{\alpha+1}}d_{p}y
\]
from equation (\ref{eq_Vl_phi}), we obtain the equation with an
absorbing region (\ref{Ab_ar}). Thus, the solution
$\phi\left(x,t\right)$ of the Cauchy problem (\ref{Ab_ar}) is the
distribution density of the stochastic process,
$\eta\left(t,\omega\right)$, whose trajectories, having reached
the domain $B_{r}\left(a\right)$, do not leave it any longer. For
the process $\eta\left(t,\omega\right)$, any passage to the domain
$B_{r}\left(a\right)$ is always the first and the only one. In
this case the distribution density of the passage time to the
domain $B_{r}\left(a\right)$ is given by

\begin{equation}
f\left(t\right)=\dfrac{\partial}{\partial t}\underset{\mathbb{Q}_{p}}{\int}\phi\left(x,t\right)d_{p}x,\label{sol}
\end{equation}
which, with regard to (\ref{Ab_ar}), yields assertion
(\ref{sol_ret}) of Theorem 4.

While the assertion of the equivalence of integral equations
(\ref{Volt_1}) and (\ref{Volt_2}) is almost obvious, the assertion
of the equivalence of equation (\ref{sol_ret}) to equations
(\ref{Volt_1}) and (\ref{Volt_2}) is not quite obvious at first
glance and requires a proof. This assertion is stated by Theorem
5.

\textbf{Theorem 5.} The solutions of equations (\ref{Volt_1}) and
(\ref{Volt_2}) can be represented in the form (\ref{sol_ret}).

\textbf{Proof.} To prove the theorem, it suffices to show that the
solutions of equations (\ref{Volt_1}) and (\ref{Volt_2}) in terms
of Laplace transforms coincide with the Laplace transform of the
function (\ref{sol_ret}). Consider for definiteness equation
(\ref{Volt_2}). Since the functions $g\left(t\right)$ and
$g_{R}\left(t\right)$ are continuous, this equation has a unique
solution in the same class of functions \cite{P_M}. The functions
$g\left(t\right)$ and $g_{R}\left(t\right)$ are bounded above;
i.e., they have zero growth exponent; therefore, we will seek a
solution $f\left(t\right)$ of equation (\ref{Volt_2}) in the class
of functions with finite growth exponent. In terms of Laplace
transforms equation (\ref{Volt_2}) has the form

\begin{equation}
G\left(s\right)=G_{R}\left(s\right)F\left(s\right)+F\left(s\right),\label{Volt_2_Lap}
\end{equation}
where $G\left(s\right)$, $G_{R}\left(s\right)$, and
$F\left(s\right)$ are the Laplace transforms of the functions
$g\left(t\right)$, $g_{R}\left(t\right)$, and $f\left(t\right)$
respectively. From (\ref{Volt_2_Lap}) we obtain

\begin{equation}
F\left(s\right)=\dfrac{G\left(s\right)}{1+G_{R}\left(s\right)}.\label{F(s)}
\end{equation}
From the definitions of $g\left(t\right)$ and
$g_{R}\left(t\right)$ (see (\ref{g(t)_T1}) and (\ref{g(t)_R_T1}))
we can easily find $G\left(s\right)$ and $G_{R}\left(s\right)$.
Let us first find $G\left(s\right)$. To this end we represent
$g\left(t\right)$ as

\[
g\left(t\right)=-\dfrac{1}{\Gamma_{p}\left(-\alpha\right)}\underset{B_{r}}{\int}d_{p}y\underset{\mathbb{Q}_{p}\setminus B_{r}}{\int}\dfrac{\varphi\left(x+a,t\right)}{\left|x-y\right|_{p}^{\alpha+1}}d_{p}x
\]
\begin{equation}
=-\dfrac{p^{r}}{\Gamma_{p}\left(-\alpha\right)}\underset{\mathbb{Q}_{p}\setminus B_{r}}{\int}\dfrac{\varphi\left(x+a,t\right)}{\left|x\right|_{p}^{\alpha+1}}d_{p}x\label{g(t)_calc}
\end{equation}
and substitute into (\ref{g(t)_calc}) the solution of the Cauchy
problem (\ref{p-KF}), (\ref{I_C}) expressed in terms of the
$p$-adic analog of the Fourier integral:

\[
\varphi\left(x+a,t\right)=\underset{\mathbb{Q}_{p}}{\int}\Omega\left(\left|k\right|_{p}\right)\exp\left(-\left|k\right|_{p}^{\alpha}t\right)\chi\left(-k\left(x+a\right)\right)d_{p}k.
\]
Then we have

\begin{equation}
g\left(t\right)=p^{r}\underset{\mathbb{Q}_{p}}{\int}\Omega\left(\left|k\right|_{p}\right)\exp\left(-\left|k\right|_{p}^{\alpha}t\right)\chi\left(-ka\right)\left(B_{\alpha}\left(r\right)-\left|k\right|_{p}^{\alpha}\right)\Omega\left(\left|k\right|_{p}p^{r}\right)d_{p}k.\label{g(t)_calc_next}
\end{equation}
Passing to the Laplace transform in (\ref{g(t)_calc_next}), we
obtain

\begin{equation}
G\left(s\right)=p^{r}\underset{\mathbb{Q}_{p}}{\int}\Omega\left(\left|k\right|_{p}\right)\Omega\left(\left|k\right|_{p}p^{r}\right)\dfrac{B_{\alpha}\left(r\right)-\left|k\right|_{p}^{\alpha}}{\left|k\right|_{p}^{\alpha}+s}\chi\left(-ka\right)d_{p}k.\label{G(s)_Laplace}
\end{equation}
Similarly we calculate $G_{R}\left(s\right)$. Changing the
variables and integrating in (\ref{g(t)_R_T1}), we obtain
\begin{equation}
g_{R}\left(t\right)=-\dfrac{p^{r}}{\Gamma_{p}\left(-\alpha\right)}\underset{\mathbb{Q}_{p}\setminus B_{r}}{\int}\dfrac{\psi\left(x+a,t\right)}{\left|x\right|_{p}^{\alpha+1}}d_{p}x.\label{g(t)_calc_R}
\end{equation}
Let us substitute into (\ref{g(t)_calc_R}) the solution of the
Cauchy problem for equation (\ref{p-KF}) with the initial
condition (\ref{NU_B}), expressed in the form of a Fourier
integral
\[
\psi\left(x+a,t\right)=\underset{\mathbb{Q}_{p}}{\int}\Omega\left(\left|k\right|_{p}p^{r}\right)\exp\left(-\left|k\right|_{p}^{\alpha}t\right)\chi\left(-kx\right)d_{p}k.
\]
Integrating the expression obtained with respect to $x$, we have
\begin{equation}
g_{R}\left(t\right)=p^{r}\underset{\mathbb{Q}_{p}}{\int}\Omega\left(\left|k\right|_{p}p^{r}\right)\exp\left(-\left|k\right|_{p}^{\alpha}t\right)\left(B_{\alpha}\left(r\right)-\left|k\right|_{p}^{\alpha}\right)d_{p}k.\label{g(t)_calc_R_1}
\end{equation}
Passing to Laplace transforms in (\ref{g(t)_calc_R_1}), we obtain

\begin{equation}
G_{R}\left(s\right)=p^{r}\underset{\mathbb{Q}_{p}}{\int}\Omega\left(\left|k\right|_{p}p^{r}\right)\dfrac{B_{\alpha}\left(r\right)-\left|k\right|_{p}^{\alpha}}{s+\left|k\right|_{p}^{\alpha}}d_{p}k.\label{G(S)_R_Laplace}
\end{equation}

Now, let us show that the Laplace transform of the function
(\ref{sol_ret}) coincides with (\ref{F(s)}). In the Cauchy problem
(\ref{p-KF}), (\ref{NU_B}) we make the change of variables
$x-a=x^{\prime}$, $y-a=y^{\prime}$ and pass to the Fourier
transforms. Then the Cauchy problem is rewritten as
\[
\dfrac{d}{dt}\tilde{\psi}\left(k,t\right)\chi\left(-ka\right)=-\chi\left(-ka\right)\left|k\right|_{p}^{\alpha}\tilde{\psi}\left(k,t\right)
\]

\begin{equation}
-p^{r}\Omega\left(\left|k\right|p^{r}\right)\underset{\mathbb{Q}_{p}}{\int}d_{p}q\Omega\left(\left|k\right|_{p}p^{r}\right)\tilde{\psi}\left(k,t\right)\left(B_{\alpha}\left(r\right)-\left|q\right|_{p}^{\alpha}\right)\chi\left(-qa\right),\label{psi_eq}
\end{equation}

\[
\tilde{\psi}\left(k,0\right)=\chi\left(-ka\right)\Omega\left(\left|k\right|_{p}\right).
\]
Let us write equation (\ref{psi_eq}) in terms of Laplace
transforms:

\[
s\tilde{\varPsi}\left(k,s\right)\chi\left(-ka\right)=\chi\left(-ka\right)\Omega\left(\left|k\right|_{p}\right)-\chi\left(-ka\right)\left|k\right|_{p}^{\alpha}\tilde{\varPsi}\left(k,s\right)-
\]

\[
-p^{r}\Omega\left(\left|k\right|_{p}p^{r}\right)\underset{\mathbb{Q}_{p}}{\int}d_{p}q\tilde{\varPsi}\left(q,s\right)\Omega\left(\left|q\right|_{p}p^{r}\right)\left(B_{\alpha}\left(r\right)-\left|q\right|_{p}^{\alpha}\right)\chi\left(-qa\right),
\]
where $\tilde{\varPsi}\left(k,s\right)$ is the Laplace transform
of the function $\tilde{\psi}\left(k,t\right)$. The last equation
can be represented as

\[
\tilde{\varPsi}\left(k,s\right)\chi\left(-ka\right)=\dfrac{\chi\left(-ka\right)\Omega\left(\left|k\right|_{p}\right)}{s+\left|k\right|_{p}^{\alpha}}-
\]

\begin{equation}
-\dfrac{\Omega\left(\left|k\right|_{p}p^{r}\right)}{s+\left|k\right|_{p}^{\alpha}}p^{r}\underset{\mathbb{Q}_{p}}{\int}d_{p}q\tilde{\varPsi}\left(q,s\right)\Omega\left(\left|q\right|_{p}p^{r}\right)\left(B_{\alpha}\left(r\right)-\left|q\right|_{p}^{\alpha}\right)\chi\left(-qa\right).\label{Psi_eq}
\end{equation}
Let us multiply (\ref{Psi_eq}) by the expression

\[
p^{r}\Omega\left(\left|k\right|_{p}p^{r}\right)\left(B_{\alpha}\left(r\right)-\left|k\right|_{p}^{\alpha}\right)
\]
and integrate the result with respect to the variable $k$ over the
field $\mathbb{Q}_{p}$. Then, using representation (\ref{sol_ret})
in terms of Laplace and Fourier transforms,
\[
F\left(s\right)=p^{r}\underset{\mathbb{Q}_{p}}{\int}d_{p}q\tilde{\varPsi}\left(q,s\right)\Omega\left(\left|q\right|_{p}p^{r}\right)\left(B_{\alpha}\left(r\right)-\left|q\right|_{p}^{\alpha}\right)\chi\left(-qa\right),
\]
we obtain
\[
F\left(s\right)=p^{r}\underset{\mathbb{Q}_{p}}{\int}d_{p}k\Omega\left(\left|k\right|_{p}\right)\Omega\left(\left|k\right|_{p}p^{r}\right)\dfrac{B_{\alpha}\left(r\right)-\left|k\right|_{p}^{\alpha}}{s+\left|k\right|_{p}^{\alpha}}\chi\left(-ka\right)
\]

\begin{equation}
-p^{r}\underset{\mathbb{Q}_{p}}{\int}d_{p}k\Omega\left(\left|k\right|_{p}p^{r}\right)\dfrac{B_{\alpha}\left(r\right)-\left|k\right|_{p}^{\alpha}}{s+\left|k\right|_{p}^{\alpha}}F\left(s\right).\label{F(s)_proof}
\end{equation}
Comparing (\ref{F(s)_proof}) with expressions (\ref{G(s)_Laplace})
and (\ref{G(S)_R_Laplace}) for $G\left(s\right)$ and
$G_{R}\left(s\right)$, we obtain (\ref{Volt_2_Lap}), which proves
Theorem 5.

\section{Properties of the distribution density $f\left(t\right)$}

Let us proceed to the analysis of the distribution density
$f\left(t\right)$. First, we represent its Laplace transform
(\ref{F(s)}) in a form convenient for subsequent analysis.
Calculating the integrals in (\ref{G(s)_Laplace}) and
(\ref{G(S)_R_Laplace}), we obtain

\begin{equation}
G_{R}\left(s\right)=-1+p^{r}\left(B_{\alpha}\left(r\right)+s\right)J_{r}\left(s\right),\label{GR}
\end{equation}

\begin{equation}
G\left(s\right)=p^{r}\left(B_{\alpha}\left(r\right)+s\right)E\left(s,\left|a\right|_{p}\right),\label{G}
\end{equation}
where

\begin{equation}
E\left(s,\left|a\right|_{p}\right)=\dfrac{1}{\left|a\right|_{p}}\left[\left(1-\dfrac{1}{p}\right)\stackrel[n=0]{\infty}{\sum}\dfrac{p^{-n}}{s+p^{-\alpha n}\left|a\right|_{p}^{-\alpha}}-\dfrac{1}{s+p^{\alpha}\left|a\right|_{p}^{-\alpha}}\right],\label{E}
\end{equation}

\begin{equation}
J_{r}\left(s\right)=\left(1-\dfrac{1}{p}\right)\stackrel[n=r]{\infty}{\sum}\dfrac{p^{-n}}{s+p^{-\alpha n}}.\label{J}
\end{equation}
Then (\ref{F(s)}) implies

\begin{equation}
F\left(s\right)=\dfrac{E\left(\left|a\right|_{p},s\right)}{J_{r}\left(s\right)}.\label{F(s)_sol}
\end{equation}
Substituting expressions (\ref{E}) and (\ref{J}) into
(\ref{F(s)_sol}), we obtain

\begin{equation}
F\left(s\right)=1-\dfrac{\left(1-\dfrac{1}{p}\right)\stackrel[n=r]{\nu-1}{\sum}\dfrac{p^{-n}}{s+p^{-\alpha n}}+\dfrac{p^{-\nu}}{s+p^{-\alpha\left(\nu-1\right)}}}{\left(1-\dfrac{1}{p}\right)\stackrel[n=r]{\infty}{\sum}\dfrac{p^{-n}}{s+p^{-\alpha n}}},\label{F(s)_sol_1}
\end{equation}
where $\left|a\right|=p^{\nu}$. Expression (\ref{F(s)_sol_1}) is
key for the analysis of the properties of the function
$f\left(t\right)$.

First of all we establish the following properties of
$f\left(t\right)$.

\textbf{Property 1.} The following equations hold:

\begin{equation}
\stackrel[0]{\infty}{\int}f\left(t\right)dt=F\left(0\right)=1\;\forall\alpha\geq1,\label{Prop_1a}
\end{equation}

\begin{equation}
\stackrel[0]{\infty}{\int}f\left(t\right)dt=F\left(0\right)=\left(\dfrac{p^{r}}{\left|a\right|_{p}}\right)^{1-\alpha}\dfrac{p^{\alpha}-1}{p^{\alpha}}\dfrac{p}{p-1}<1\;\forall\:0<\alpha<1.\label{Prop_1b}
\end{equation}

Indeed, it follows from the properties of integration of the
original and passage to the limit as $s\rightarrow0$ in the base
$\mathrm{Re}s>0$ that

\[
\stackrel[0]{\infty}{\int}f\left(t\right)dt=\underset{t\rightarrow\infty}{\lim}\stackrel[0]{t}{\int}f\left(\tau\right)d\tau=\underset{s\rightarrow0}{\lim}s\left(\dfrac{F\left(s\right)}{s}\right)=\underset{s\rightarrow0}{\lim}F\left(s\right),
\]
whence, taking into account (\ref{F(s)_sol_1}), we obtain
properties (\ref{Prop_1a}) and (\ref{Prop_1b}). Property
(\ref{Prop_1a}) implies that for $\alpha\geq1$ the trajectory of
the stochastic process $\xi\left(t,\omega\right)$ certainly
reaches the domain $B_{r}\left(a\right)$ and, hence, returns there
infinitely many times in infinite time. Property (\ref{Prop_1b})
implies that for $\alpha<1$ the full probability measure of the
event of passage of the trajectory to the domain
$B_{r}\left(a\right)$ is less than one. This means that there
exist trajectories of the stochastic process
$\xi\left(t,\omega\right)$ that never reach the domain
$B_{r}\left(a\right)$. Notice that a property analogous to
property (\ref{Prop_1b}) for the distribution density
$f_{ret}\left(t\right)$, established in \cite{ABZ_2009}, has the
form

\[
\stackrel[0]{\infty}{\int}f_{ret}\left(t\right)dt=F_{ret}\left(0\right)=\left(\dfrac{p^{\alpha}-1}{p-1}\right)^{2}\dfrac{p}{p^{\alpha}}<1,\;0<\alpha<1.
\]

\textbf{Property 2. } For $\alpha\geq1$ there holds

\begin{equation}
\left\langle \tau_{B_{r}\left(a\right)}\right\rangle =\lim_{T\rightarrow\infty}\stackrel[0]{T}{\int}tf\left(t\right)dt=\infty.\label{Prop_2}
\end{equation}
Indeed, it follows from the properties of the Laplace
transformation that $-\dfrac{d}{ds}F\left(s\right)$ is the Laplace
transform of $tf\left(t\right)$, and then

\[
\underset{t\rightarrow\infty}{\lim}\stackrel[0]{t}{\int}\tau f\left(\tau\right)d\tau\risingdotseq-\underset{s\rightarrow0}{\lim}s\left(\dfrac{1}{s}\dfrac{d}{ds}F\left(s\right)\right)=-\underset{s\rightarrow0}{\lim}\dfrac{d}{ds}F\left(s\right)=\infty,
\]
where the passage to the limit is performed in the base
$\mathrm{Re}s>0$.

\textbf{Property 3. } The following holds:

\begin{equation}
\underset{t\rightarrow0+0}{\lim}f\left(t\right)=-\dfrac{1}{\Gamma_{p}\left(-\alpha\right)}\dfrac{p^{r}}{\left|a\right|_{p}^{\alpha+1}},\label{Prop_3a}
\end{equation}

\begin{equation}
\underset{t\rightarrow\infty}{\lim}f\left(t\right)=0.\label{Prop_3b}
\end{equation}
Indeed, by the initial value theorem, we have

\[
\underset{t\rightarrow0+0}{\lim}f\left(t\right)=\underset{s\rightarrow\infty}{\lim}sF\left(s\right)=-\dfrac{1}{\Gamma_{p}\left(-\alpha\right)}\dfrac{p^{r}}{\left|a\right|_{p}^{\alpha+1}},
\]

\[
\underset{t\rightarrow\infty}{\lim}f\left(t\right)=\underset{s\rightarrow0}{\lim}sF\left(s\right)=0,\;\mathrm{Re}s>0.
\]
Note for comparison that the behavior of the distribution density
function $f_{ret}\left(t\right)$ at zero (see \cite{ABZ_2009}) is
different from the behavior of the function $f\left(t\right)$:

\[
\underset{t\rightarrow0+0}{\lim}f_{ret}\left(t\right)=0.
\]

\textbf{Property 4.} The behavior of the derivatives of
$f\left(t\right)$ at zero and at infinity is as follows:

\begin{equation}
\underset{t\rightarrow\infty}{\lim}\dfrac{d}{dt}f\left(t\right)=0,\label{Prop_4a}
\end{equation}

\begin{equation}
\underset{t\rightarrow0+0}{\lim}\dfrac{d}{dt}f\left(t\right)=p^{r}\left(p^{\alpha}-1\right)\left(\dfrac{p}{\left|a\right|_{p}}\right)^{2\alpha+1}\left(\dfrac{p-1}{\left(p^{\alpha+1}-1\right)^{2}}\left(\dfrac{\left|a\right|_{p}}{p^{r}}\right)^{\alpha}-\dfrac{p^{\alpha}+1}{p^{2\alpha+1}-1}\right).\label{Prop_4b}
\end{equation}
Property (\ref{Prop_4a}) follows from

\[
\underset{t\rightarrow\infty}{\lim}\dfrac{d}{dt}f\left(t\right)\risingdotseq\underset{s\rightarrow0}{\lim}s\left(sF\left(s\right)-f\left(0\right)\right)=0\;\mathrm{Re}s>0.
\]
To prove property (\ref{Prop_4b}), consider

\[
\underset{t\rightarrow0+0}{\lim}\dfrac{d}{dt}f\left(t\right)=\underset{s\rightarrow\infty}{\lim}s\left(sF\left(s\right)-f\left(0\right)\right)=\underset{s\rightarrow\infty}{\lim}\dfrac{s^{2}E\left(\left|a\right|_{p},s\right)-sf\left(0\right)J_{r}\left(s\right)}{J_{r}\left(s\right)}.
\]
Then, expanding the functions $E\left(\left|a\right|_{p},s\right)$
and $J_{r}\left(s\right)$ in the small parameter
$\varepsilon=\dfrac{1}{s}$ up to the second order inclusive, we
obtain (\ref{Prop_4b}).

From (\ref{Prop_4b}) we easily see that

\begin{equation}
\underset{t\rightarrow0+0}{\lim}\dfrac{d}{dt}f\left(t\right)<0,\;\left|a\right|_{p}=p^{r+1},\label{d_f<0}
\end{equation}

\begin{equation}
\underset{t\rightarrow0+0}{\lim}\dfrac{d}{dt}f\left(t\right)>0\;\left|a\right|_{p}>p^{r+1}.\label{d_f>0}
\end{equation}
For comparison we also note that the derivative of the probability
density function $f_{ret}\left(t\right)$ at zero is always
positive:

\[
\underset{t\rightarrow0+0}{\lim}\dfrac{d}{dt}f_{ret}\left(t\right)=\left(p-1\right)\dfrac{p^{2\alpha+1}\left(p^{\alpha}-1\right)^{2}}{\left(p^{2\alpha+1}-1\right)\left(p^{\alpha+1}-1\right)^{2}}>0.
\]

\textbf{Property 5.} The function $f\left(t\right)$ is the same
for any initial condition of the form
$\varphi\left(x,0\right)=p^{-\sigma}\Omega\left(\left|x\right|p^{\sigma}\right)$
for the process $\xi\left(t,\omega\right)$ and depends only on the
parameters $\left|a\right|_{p}$ and $r$ of the domain
$B_{r}\left(a\right)$. This property is proved by direct
calculation of the functions $G\left(s\right)$ and
$G_{R}\left(s\right)$ for the Cauchy problem equation (\ref{p-KF})
with the initial condition (\ref{I_C}).

It follows from (\ref{F(s)_sol_1} ) that the numerator of the
function $F\left(s\right)$ is a holomorphic function on the whole
complex plane except a finite number of simple poles. The
denominator of the function $F\left(s\right)$ is a holomorphic
function on the whole complex plane except an infinite number of
simple poles, and zero is an accumulation point of the poles.
Henceforth it will be convenient to consider the function
$F\left(s\right)$ on the whole complex plane. To this end we
define $F\left(s\right)$ at the singular points $s=p^{-\alpha n}$
for any $n\geq r$ by the limits at these points:

\[
\underset{s\rightarrow-p^{-\alpha n}}{\lim}F\left(s\right)=\begin{cases}
0, & r\leq n<\nu-1\\
-\dfrac{1}{p-1}, & n=\nu-1\\
1, & n>\nu-1.
\end{cases}
\]
Then the function $F\left(s\right)$ will be holomorphic on the
whole complex plane except for the points at which it has simple
poles, $s=-\lambda_{k},\;k=0,1,2,...$ , whose existence is
established by graphical analysis of the equation
$J_{r}\left(s\right)=0$. It is easy to establish from this
analysis that the values of $\lambda_{k}$ lie in the interval
$p^{-\alpha\left(r+k+1\right)}<\lambda_{k}<p^{-\alpha\left(r+k\right)}$.
The point $s=0$ is an essentially singular point (an accumulation
point of poles). Thus, the function $F\left(s\right)$ is not a
meromorphic function on the whole complex plane, which complicates
the application of residue theory to the inverse Laplace
transformation. Nevertheless, the following theorem is valid.

\textbf{Theorem 6.} The function $F\left(s\right)$ can be
represented as an infinite sum of simple poles
\begin{equation}
F\left(s\right)=\stackrel[k=0]{\infty}{\sum}\dfrac{b_{k}}{s+\lambda_{k}},\label{F(s)_Th_6}
\end{equation}
where $b_{k}$ are the residues of $F\left(s\right)$ at the poles
$-\lambda_{k},\;k=0,1,2,...$, the series (\ref{F(s)_Th_6})
converging uniformly in $s$ on any compact set of the complex
plane except for the simple poles $s=-\lambda_{k}$, and in the
half-plane $\mathrm{Re}s\geq0$.

\textbf{Proof. } Introduce an auxiliary function
$\Phi\left(z\right)=F\left(\dfrac{1}{z}\right)$, which is regular
at all points of the complex plane except for the simple poles
$z_{k}=-\dfrac{1}{\lambda_{k}},\;k=0,1,2,...$, and thus is
meromorphic. By the Mittag--Leffler theorem, any meromorphic
function can be expanded in a series

\[
\Phi\left(z\right)=h\left(z\right)+\stackrel[n=0]{\infty}{\sum}\left(g_{n}\left(z\right)-P_{n}\left(z\right)\right)
\]
that converges uniformly on any compact set, where
$h\left(z\right)$ is an entire function, $g_{n}\left(z\right)$ are
the principal parts of $\Phi\left(z\right)$, and
$P_{n}\left(z\right)$ are polynomials. Then, since
$\underset{z\rightarrow\infty}{\lim}\left|\Phi\left(z\right)\right|=F\left(0\right)$,
the following representation is valid for the function
$\Phi\left(z\right)$ in the form of a uniformly converging series
on any compact set except for the poles $z_{k}$:

\[
\Phi\left(z\right)=\stackrel[k=0]{\infty}{\sum}\left(\dfrac{c_{k}}{\dfrac{1}{s}+\dfrac{1}{\lambda_{k}}}-p_{k}\right)+c,
\]
where $p_{k}$ and $c$ are constants and $c_{k}$ are the residues
of $\Phi\left(z\right)$ at the poles $z_{k}$. Thus, we have

\[
F\left(s\right)=\stackrel[k=0]{\infty}{\sum}\left(\dfrac{c_{k}}{\dfrac{1}{s}+\dfrac{1}{\lambda_{k}}}-p_{k}\right)+c=\stackrel[k=0]{\infty}{\sum}\left(\dfrac{c_{k}\lambda_{k}s}{s+\lambda_{k}}-p_{k}\right)+c
\]
\[
=\stackrel[k=0]{\infty}{\sum}\left(-\dfrac{c_{k}\lambda_{k}^{2}}{s+\lambda_{k}}+c_{k}\lambda_{k}-p_{k}\right)+c.
\]
Since, as easily seen,
$\underset{z\rightarrow0}{\lim}\Phi\left(z\right)$=$\underset{s\rightarrow\infty}{\lim}F\left(s\right)=0$,
we have

\[
F\left(s\right)=\stackrel[k=0]{\infty}{\sum}\dfrac{\left(-c_{k}\lambda_{k}^{2}\right)}{s+\lambda_{k}}.
\]
Denoting $-c_{k}\lambda_{k}^{2}=b_{k}$, we obtain
(\ref{F(s)_Th_6}).

The uniform convergence of (\ref{F(s)_Th_6}) is obvious. Indeed,
$\forall$$\mathrm{Re}s\geq0$ there holds
$\left|\dfrac{b_{k}}{s+\lambda_{k}}\right|\leq\dfrac{\left|b_{k}\right|}{\lambda_{k}}$,
and from properties 1 and 2 we find that
$F\left(0\right)=const>0\;\forall\alpha>0$; then (\ref{F(s)_Th_6})
implies that
$\stackrel[k=0]{\infty}{\sum}\dfrac{b_{k}}{\lambda_{k}}$ is a
converging number series, thus proving the uniform convergence.
Theorem 6 is proved.

Since the series (\ref{F(s)_Th_6}) is uniformly continuous, we can
easily obtain an expression for the original $f\left(t\right)$ by
applying the inverse Laplace transformation to each term of the
series:

\[
f\left(t\right)=L_{s\rightarrow t}^{-1}\left[F\left(s\right)\right]\left(t\right)=\stackrel[k=0]{\infty}{\sum}b_{k}L_{s\rightarrow t}^{-1}\left[\dfrac{1}{s+\lambda_{k}}\right]\left(t\right)
\]
or

\begin{equation}
f\left(t\right)=\stackrel[k=0]{\infty}{\sum}b_{k}\exp\left(-\lambda_{k}t\right),\label{f(t)_sol}
\end{equation}
where the coefficients $b_{k}$ are the residues of the function
$F\left(s\right)$ at simple poles $s=-\lambda_{k}$. To find
$b_{k}$, we write
\[
b_{k}=\underset{s=-\lambda_{k}}{\mathrm{res}}\left(F\left(s\right)\right)=\underset{s\rightarrow\lambda_{k}}{\lim}\left(s+\lambda_{k}\right)\dfrac{E\left(\left|a\right|_{p},s\right)}{J_{r}\left(s\right)}
\]

\[
=E\left(\left|a\right|_{p},-\lambda_{k}\right)\underset{s\rightarrow-\lambda_{k}}{\lim}\dfrac{s+\lambda_{k}}{J_{r}\left(s\right)}=E\left(\left|a\right|_{p},-\lambda_{k}\right)\underset{s\rightarrow-\lambda_{k}}{\lim}\dfrac{1}{\dfrac{d}{ds}J_{r}\left(s\right)}.
\]
Let us calculate

\[
\dfrac{d}{ds}J_{r}\left(s\right)=\left(1-p^{-1}\right)\dfrac{d}{ds}\stackrel[n=r]{\infty}{\sum}\dfrac{p^{-n}}{s+p^{-\alpha n}}
\]

\[
=\left(1-p^{-1}\right)\dfrac{d}{ds}\dfrac{1}{s}\stackrel[n=r]{\infty}{\sum}\dfrac{p^{-n}}{1+\dfrac{p^{-\alpha n}}{s}}.
\]
The sum in the last expression is uniformly convergent by the Abel
test; after differentiation we obtain

\[
\dfrac{d}{ds}J_{r}\left(s\right)=-\left(1-p^{-1}\right)\stackrel[n=r]{\infty}{\sum}\dfrac{p^{-n}}{\left(p^{-\alpha n}-s\right)^{2}},
\]
whence we have

\begin{equation}
b_{k}=\dfrac{\left(1-\dfrac{1}{p}\right)\stackrel[n=r]{\nu-1}{\sum}\dfrac{p^{-n}}{p^{-\alpha n}-\lambda_{k}}+\dfrac{p^{-\nu}}{p^{-\alpha\left(\nu-1\right)}-\lambda_{k}}}{\left(1-\dfrac{1}{p}\right)\stackrel[n=r]{\infty}{\sum}\dfrac{p^{-n}}{\left(p^{-\alpha n}-\lambda_{k}\right)^{2}}}.\label{coef_b}
\end{equation}

Let us show that the series (\ref{f(t)_sol}) is uniformly
convergent. By property 4,
$\stackrel[k=0]{\infty}{\sum}b_{k}=\dfrac{p^{\alpha}-1}{1-p^{-\left(\alpha+1\right)}}\dfrac{p^{r}}{\left|a\right|_{p}^{\alpha+1}}$.
Next,
$\left|\exp\left(-\lambda_{k}t\right)\right|\leq1\;\forall\,t\in\mathbb{R}_{+}$
and $\forall\,k=0,1,2,\ldots$; then, by the Abel test, the series
$\stackrel[k=0]{\infty}{\sum}b_{k}\exp\left(-\lambda_{k}t\right)$
converges uniformly and $f\left(t\right)$ is a continuous
function. These results can be formulated as the following
theorem.

\textbf{Theorem 7.} The probability density function
$f\left(t\right)$ of the random variable
$\tau_{B_{r}\left(a\right)}\left(\omega\right)$ is represented as
a uniformly convergent series (\ref{f(t)_sol}) with coefficients
defined by (\ref{coef_b}).

The following theorem holds.

\textbf{Theorem 8.} The function $f\left(t\right)$ is positive, it
monotonically decreases for $\left|a\right|_{p}=p^{r+1}$, and has
a unique maximum at point $t=t_{\max}>0$ for
$\left|a\right|>p^{r+1}$.

\textbf{Proof.} If $\left|a\right|=p^{r+1}$, setting $\nu=r+1$, we
can easily see from (\ref{coef_b}) that
$b_{k}>0\;\forall\:k\geq0$, which implies positivity, monotonic
decrease, and strict downward convexity. If
$\left|a\right|>p^{r+1}$, then, in view of property (\ref{d_f>0})
we have

\[
\underset{t\rightarrow0+0}{\lim}\dfrac{d}{dt}f\left(t\right)>0,
\]
whence, taking into account (\ref{f(t)_sol}), we obtain

\[
-\stackrel[k=0]{\infty}{\sum}\lambda_{k}b_{k}>0.
\]
Since $\lambda_{k}>0$, it follows that $b_{k}<0$ for the first $m$
terms, where $m<\nu-r-1$, while, for $k>m$, all $b_{k}>0$. For
example, one can show from (\ref{coef_b}) by direct calculation
that, for $\nu=r+2$, $b_{0}<0$ and $b_{k}>0\;\forall\,k=1,2,...$,
that is, $m=0$. For $\nu=r+3$, also $b_{0}<0$ and
$b_{k}>0\;\forall\,k=1,2,...$, and $m=0$. Let us show that
(\ref{f(t)_sol}) is a positive function on $\mathbb{R}_{+}$. To
this end, we rewrite it as

\[
f\left(t\right)=\stackrel[k=0]{\infty}{\sum}b_{k}\exp\left(-\lambda_{k}t\right)=f_{+}\left(t\right)-f_{-}\left(t\right)
\]

\[
=\left|b_{m}\right|\exp\left(-\lambda_{m}t\right)\left(-\dfrac{1}{\left|b_{m}\right|}\stackrel[k=0]{m-1}{\sum}\left|b_{k}\right|\exp\left(-\left(\lambda_{k}-\lambda_{m}\right)t\right)-1+\dfrac{1}{\left|b_{m}\right|}\stackrel[k=m+1]{\infty}{\sum}b_{k}\exp\left(\left(\lambda_{m}-\lambda_{k}\right)t\right)\right),
\]
where

\[
f_{-}\left(t\right)=\stackrel[k=0]{m}{\sum}\left|b_{k}\right|\exp\left(-\lambda_{k}t\right),\;f_{+}\left(t\right)=\stackrel[k=m+1]{\infty}{\sum}b_{k}\exp\left(-\lambda_{k}t\right).
\]
By (\ref{Prop_3a}),

\[
-\dfrac{1}{\left|b_{m}\right|}\stackrel[k=0]{m-1}{\sum}\left|b_{k}\right|-1+\dfrac{1}{\left|b_{m}\right|}\stackrel[k=m+1]{\infty}{\sum}b_{k}>0,
\]
whence it follows that $f\left(t\right)>0$ for all $t>0$.

Let us show that $f\left(t\right)$ has a maximum at some point
$t=t_{\max}$ and this maximum is unique. Consider the derivative
of $f\left(t\right)$,

\[
\dfrac{d}{dt}f\left(t\right)=\stackrel[k=0]{\infty}{-\sum}\lambda_{k}b_{k}\exp\left(-\lambda_{k}t\right)=\stackrel[k=0]{m}{\sum}\lambda_{k}\left|b_{k}\right|\exp\left(-\lambda_{k}t\right)-\stackrel[k=m+1]{\infty}{\sum}\lambda_{k}b_{k}\exp\left(-\lambda_{k}t\right)
\]
\[
=\lambda_{m}\left|b_{m}\right|\exp\left(-\lambda_{k}t\right)\left(g\left(t\right)-h\left(t\right)\right),
\]
where
\[
g\left(t\right)=\dfrac{1}{\lambda_{m}\left|b_{m}\right|}\stackrel[k=0]{m-1}{\sum}\lambda_{k}\left|b_{k}\right|\exp\left(-\left(\lambda_{k}-\lambda_{m}\right)t\right)+1,
\]
\[
h\left(t\right)=\dfrac{1}{\lambda_{m}\left|b_{m}\right|}\stackrel[k=m+1]{\infty}{\sum}\lambda_{k}b_{k}\exp\left(\left(\lambda_{m}-\lambda_{k}\right)t\right).
\]
The function $g\left(t\right)$ monotonically decreases, while
$h\left(t\right)$ monotonically increases. From (\ref{Prop_4b}) we
have
$\left|\dfrac{d}{dt}f_{-}\left(t\right)\mid_{t=0}\right|>\left|\dfrac{d}{dt}f_{+}\left(t\right)\mid_{t=0}\right|$,
whence $g\left(0\right)>h\left(0\right)$. Then
$\exists\;t=t_{\max}\colon\:g\left(t_{\max}\right)=h\left(t_{\max}\right)$.
Hence it follows that
$\dfrac{d}{dt}f\left(t\right)\mid_{t=t_{\max}}=0$ and
$\dfrac{d}{dt}f\left(t\right)>0$ if $t<t_{\max}$, and
$\dfrac{d}{dt}f\left(t\right)<0$ if $t>t_{\max}$. This completes
the proof of Theorem 8.

For comparison, note that the function $f_{ret}\left(t\right)$
exhibits similar behavior. As already pointed out above,
$f_{ret}\left(0\right)=0$; then, for $t>0$, it increases until it
reaches a maximum at some $t=t_{\max}$, and then, for
$t>t_{\max}$, it decreases to zero.

The asymptotic behavior of $f\left(t\right)$ for large times is
established by the following theorem.

\textbf{Theorem 9.} The function $f\left(t\right)$ defined by the
series (\ref{f(t)_sol}) exhibits the following asymptotic behavior
as $t\rightarrow\infty:$
\[
f\left(t\right)=I_{0}^{\nu-r-1}\left(0\right)\dfrac{\left(p^{-\alpha}+\Delta\left(\alpha\right)\right)^{-\tfrac{\left(2\alpha-1\right)}{\alpha}}}{\stackrel[n=-\infty]{\infty}{\sum}\dfrac{p^{n}}{\left(p^{\alpha n}-p^{-\alpha}-\Delta\left(\alpha\right)\right)^{2}}}\dfrac{p^{\left(\alpha-1\right)r}}{\alpha\ln p}
\]

\begin{equation}
\times t^{-\tfrac{2\alpha-1}{\alpha}}\stackrel[m=-\infty]{\infty}{\sum}\exp\left(\dfrac{2\pi im}{\alpha\ln p}\ln\dfrac{p^{-\alpha}+\Delta\left(\alpha\right)}{p^{\alpha r}}t\right)\Gamma\left(\dfrac{2\alpha-1}{\alpha}-\dfrac{2\pi im}{\alpha\ln p}\right)\left(1+o\left(t\right)\right),\;\alpha>1,\label{as_a>1}
\end{equation}
\[
f\left(t\right)=I_{0}^{\nu-r-1}\left(0\right)p^{2}\ln p
\]

\begin{equation}
\times\dfrac{t^{-1}}{\ln^{2}\left(p^{-\left(r+1\right)}t\right)^{2}}\stackrel[m=-\infty]{\infty}{\sum}\exp\left(\dfrac{2\pi im}{\ln p}\ln p^{-\left(r+1\right)}t\right)\Gamma\left(1-\dfrac{2\pi im}{\ln p}\right)\left(1+o\left(t\right)\right),\;\alpha=1,\label{as_a=00003D1}
\end{equation}
\[
f\left(t\right)=I_{0}^{\nu-r-1}\left(0\right)p^{2}\dfrac{p^{\left(1-\alpha\right)r}}{\alpha\ln p}
\]

\begin{equation}
\times t^{-\tfrac{1}{\alpha}}\stackrel[m=-\infty]{\infty}{\sum}\exp\left(\dfrac{2\pi im}{\alpha\ln p}\ln\left(\dfrac{p^{-\alpha}}{p^{\alpha r}}t\right)\right)\Gamma\left(\dfrac{1}{\alpha}-\dfrac{2\pi im}{\alpha\ln p}\right)\left(1+o\left(t\right)\right),\;0<\alpha<1,\label{as_a<1}
\end{equation}
where $p^{-\alpha
r}I_{0}^{\nu-r-1}\left(0\right)=\underset{k\rightarrow\infty}{\lim}I_{r}^{\nu-1}\left(\lambda_{k}\right)=p^{-\alpha
r}$$\left(\left(1-\dfrac{1}{p}\right)\stackrel[n=0]{\nu-r-1}{\sum}p^{\left(\alpha-1\right)n}+p^{\left(\alpha-1\right)\left(\nu-r\right)}p^{-\alpha}\right)$
and $o\left(t\right)$ is an infinitesimal function as
$t\rightarrow\infty$.

\textbf{Proof of Theorem 9.} To prove this theorem, it is
necessary to first estimate the coefficients $b_{k}$ of the
series. To this end we rewrite formula (\ref{coef_b}) in a more
compact form

\begin{equation}
b_{k}=\dfrac{I_{r}^{\nu-1}\left(-\lambda_{k}\right)}{\left|J_{r}^{\prime}\left(-\lambda_{k}\right)\right|}.\label{coef_b_compress}
\end{equation}
It is obvious that we should first estimate the values of
$\lambda_{k}$. To this end, it is convenient to represent
$\lambda_{k}$ in terms of the variable
$\Delta_{k}\left(\alpha\right)$, which satisfies the condition
$0<\Delta_{k}\left(\alpha\right)<1$:

\[
\lambda_{k}=p^{-\alpha r}p^{-\alpha k}\left(p^{-\alpha}+\Delta_{k}\left(\alpha\right)\right).
\]
Thus, we have to estimate $\Delta_{k}\left(\alpha\right)$. To this
end, we prove Lemma 1.

\textbf{Lemma 1.} For $\alpha>0$, the following general estimate
holds for $\Delta_{k}\left(\alpha\right)$:

\begin{equation}
\dfrac{1}{1+pa_{k}}<\Delta_{k}\left(\alpha\right)<\dfrac{1}{\left(p-1\right)a_{k}},\label{Eq_lemma_1}
\end{equation}
where

\begin{equation}
a_{k}=\stackrel[i=0]{k}{\sum}p^{-\left(\alpha-1\right)i}=\dfrac{1-p^{-\left(\alpha-1\right)\left(k+1\right)}}{1-p^{-\alpha+1}}.\label{a_k}
\end{equation}
In particular, the following estimates are valid:

\begin{equation}
\dfrac{1}{p+1}\dfrac{p^{\alpha-1}-1}{p^{\alpha-1}}\dfrac{p^{\alpha}-1}{p^{\alpha}}<\Delta_{k}\left(\alpha\right)<\dfrac{1}{p-1},\;\alpha>1,\label{Delta_alpha>1}
\end{equation}

\begin{equation}
\dfrac{1}{p+1}\dfrac{p^{\alpha-1}-1}{p^{\alpha-1}}\dfrac{p^{\alpha}-1}{p^{\alpha}}p^{-\left(1-\alpha\right)k}<\Delta_{k}\left(\alpha\right)<\dfrac{1}{p-1}p^{-\left(1-\alpha\right)k},\;0<\alpha<1,\label{Delta_alpha<1}
\end{equation}

\begin{equation}
\dfrac{1}{p+1}\dfrac{p^{\alpha}-1}{p^{\alpha}}\left(k+1\right)^{-1}<\Delta_{k}\left(\alpha\right)<\dfrac{p}{\left(p-1\right)^{2}}\dfrac{p^{\alpha}-1}{p^{\alpha}}\left(k+1\right)^{-1},\;\alpha=1.\label{Delta_alpha=00003D1}
\end{equation}
\textbf{Proof of Lemma 1.} From the equation
$J_{r}\left(-\lambda_{k}\right)=0$ we have
\[
\stackrel[n=0]{\infty}{\sum}\dfrac{p^{-n}}{p^{-\alpha n}-p^{-\alpha\left(k+1\right)}-p^{-\alpha k}\Delta_{k}\left(\alpha\right)}=0.
\]
Taking the term with the number $k+1$ in this sum,
\[
\dfrac{p^{-k}p^{-1}}{p^{-\alpha k}\Delta_{k}\left(\alpha\right)}=\stackrel[n=0]{k}{\sum}\dfrac{p^{-n}}{p^{-\alpha n}-p^{-\alpha\left(k+1\right)}-p^{-\alpha k}\Delta_{k}\left(\alpha\right)}
\]

\[
-\stackrel[n=k+2]{\infty}{\sum}\dfrac{p^{-k}}{p^{-\alpha\left(k+1\right)}+p^{-\alpha k}\Delta_{k}\left(\alpha\right)-p^{-\alpha n}},
\]

\[
\dfrac{p^{-k}p^{-1}}{p^{-\alpha k}\Delta_{k}\left(\alpha\right)}=\stackrel[n=0]{k}{\sum}\dfrac{p^{-n}p^{k}}{p^{-\alpha n}p^{k}-p^{-\alpha\left(k+1\right)}p^{k}-p^{-\alpha\left(k-1\right)}\Delta_{k}\left(\alpha\right)}
\]
\[
-\stackrel[n=k+2]{\infty}{\sum}\dfrac{1}{p^{-\alpha\left(k+1\right)}p^{k}+p^{-\alpha\left(k-1\right)}\Delta_{k}\left(\alpha\right)-p^{k}p^{-\alpha n}},
\]
multiplying the equality by $\dfrac{p^{k}}{p^{\alpha k}}$, and
passing to new indices $i$ and $j$, we obtain

\begin{equation}
\dfrac{p^{-1}}{\Delta_{k}\left(\alpha\right)}=\stackrel[i=0]{k}{\sum}\dfrac{p^{-\left(\alpha-1\right)i}}{1-p^{-\alpha}p^{-\alpha i}-p^{-\alpha i}\Delta_{k}\left(\alpha\right)}-\stackrel[j=2]{\infty}{\sum}\dfrac{p^{-j}}{\Delta_{k}\left(\alpha\right)+p^{-\alpha}-p^{-\alpha j}}.\label{eq_delta}
\end{equation}
Let us estimate $\Delta_{k}\left(\alpha\right)$ from below. For
the denominator of the first sum in (\ref{eq_delta}) we have
$1-p^{-\alpha}p^{-\alpha i}-p^{-\alpha
i}\Delta_{k}\left(\alpha\right)>1-p^{-\alpha}-\Delta_{k}\left(\alpha\right)$.
Then, neglecting the second sum in (\ref{eq_delta} ) in view of
its positivity, we obtain the following inequality:

\[
\dfrac{p^{-1}}{\Delta_{k}\left(\alpha\right)}<\stackrel[i=0]{k}{\sum}\dfrac{p^{-\left(\alpha-1\right)i}}{1-p^{-\alpha}-\Delta_{k}\left(\alpha\right)}.
\]
From the last inequality we can easily derive the lower estimate

\begin{equation}
\Delta_{k}\left(\alpha\right)>\dfrac{1-p^{-\alpha}}{1+pa_{k}},\label{inf_delta}
\end{equation}
whence, taking into account (\ref{a_k}), we obtain estimates
(\ref{Delta_alpha>1}) -- (\ref{Delta_alpha=00003D1}).

Let us obtain a lower estimate for
$\Delta_{k}\left(\alpha\right)$. From equation (\ref{eq_delta}) we
have  $1-p^{-\alpha}p^{-\alpha i}-p^{-\alpha
i}\left(1-p^{-\alpha}\right)\delta_{k}<1$ for the denominator of
the first sum, while, for the denominator of the second sum we
have $\Delta_{k}\left(\alpha\right)+p^{-\alpha}-p^{-\alpha
j}>\Delta_{k}\left(\alpha\right)$. Substituting these estimates
into (\ref{eq_delta}), we obtain the inequality

\[
\dfrac{p^{-1}}{\Delta_{k}\left(\alpha\right)}>\stackrel[i=0]{k}{\sum}\dfrac{p^{-\left(\alpha-1\right)i}}{1}-\stackrel[j=2]{\infty}{\sum}\dfrac{p^{-j}}{\Delta_{k}\left(\alpha\right)}.
\]
From this inequality we can easily obtain

\begin{equation}
\Delta_{k}\left(\alpha\right)<\dfrac{1}{\left(p-1\right)a_{k}},\label{sup_delta}
\end{equation}
whence, taking into account (\ref{a_k}), we deduce the lower
estimates (\ref{Delta_alpha>1})--(\ref{Delta_alpha=00003D1}).
Lemma 1 is proved.

To estimate the coefficients $b_{k}$, it is convenient to
represent them as

\begin{equation}
b_{k}=p^{-\alpha r}\dfrac{\left(1-p^{-1}\right)\stackrel[n=0]{\nu-r-1}{\sum}\dfrac{p^{-n}}{p^{-\alpha n}-\mu_{k}}+\dfrac{p^{-\left(\nu-r\right)}}{p^{-\alpha\left(\nu-r-1\right)}-\mu_{k}}}{\left(1-p^{-1}\right)\stackrel[n=0]{\infty}{\sum}\dfrac{p^{-\alpha n}}{\left(p^{-\alpha n}-\mu_{k}\right)^{2}}}=p^{-\alpha r}\dfrac{I_{0}^{\nu-r-1}\left(-\mu_{k}\right)}{\left|J_{0}^{\prime}\left(-\mu_{k}\right)\right|}.\label{coef_b_comp}
\end{equation}
To estimate the denominator in (\ref{coef_b_comp}), we prove a few
more lemmas.

\textbf{Lemma 2. } The following estimate holds for the
denominator in (\ref{coef_b_comp}):

\[
C_{1}\left(\alpha\right)p^{\left(2\alpha-1\right)k}<\dfrac{p^{-n}}{\left(p^{-\alpha n}-\mu_{k}\right)^{2}}<C_{2}\left(\alpha\right)p^{\left(2\alpha-1\right)k},\;\alpha>1,
\]
\[
C_{3}\left(\alpha\right)p^{k}<\dfrac{p^{-n}}{\left(p^{-\alpha n}-\mu_{k}\right)^{2}}<C_{4}\left(\alpha\right)p^{k},\;0<\alpha<1,
\]
\[
C_{5}\left(k+1\right)^{2}p^{k}<\dfrac{p^{-n}}{\left(p^{-\alpha n}-\mu_{k}\right)^{2}}<C_{6}\left(k+1\right)^{2}p^{k},\;\alpha=1,
\]
where
\[
C_{1}\left(\alpha\right)=\dfrac{\left(p+1\right)^{2}}{\left(1-p^{-\alpha}\right)^{2}\left(1-p^{-\left(\alpha-1\right)}\right)},
\]
\[
C_{2}\left(\alpha\right)=\dfrac{\left(p+1\right)^{2}}{\left(1-p^{-\alpha}\right)^{2}\left(1-p^{-\left(\alpha-1\right)}\right)^{2}}\left(\dfrac{1}{p-1}+\dfrac{p^{2\alpha-1}}{p^{2\alpha-1}-1}\left(\dfrac{p^{\alpha}}{\left(p-1\right)\left(p^{\alpha}-1\right)-p^{\alpha}}\right)^{2}\right),
\]
\[
C_{3}\left(\alpha\right)=\dfrac{\left(p-1\right)^{2}}{p},
\]
\[
C_{4}\left(\alpha\right)=\dfrac{\left(p+1\right)^{2}}{\left(1-p^{-\alpha}\right)^{2}\left(1-p^{-\left(1-\alpha\right)}\right)^{2}}\left(\dfrac{1}{p-1}+\left(\dfrac{\left(p-1\right)\left(p^{\alpha}-1\right)}{\left(p-1\right)\left(p^{\alpha}-1\right)-p^{\alpha}}\right)^{2}\right),
\]
\[
C_{5}=\left(p-1\right)^{2},
\]
\[
C_{6}=\left(\dfrac{p+1}{p-1}\right)^{2}p^{2}\left(\dfrac{1}{p-1}+\left(\dfrac{p}{\left(p-1\right)^{2}-p}\right)^{2}\right).
\]

\textbf{Proof of Lemma 2.} Let us get the upper estimate. We write

\[
\stackrel[n=0]{\infty}{\sum}\dfrac{p^{-n}}{\left(p^{-\alpha n}-\mu_{k}\right)^{2}}=\stackrel[n=0]{k}{\sum}\dfrac{p^{-n}}{\left(p^{-\alpha n}-p^{-\alpha\left(k+1\right)}-p^{-\alpha k}\Delta_{k}\left(\alpha\right)\right)^{2}}
\]
\[
+\dfrac{p^{-\left(k+1\right)}}{p^{-2\alpha k}\Delta_{k}^{2}\left(\alpha\right)}+\stackrel[n=k+2]{\infty}{\sum}\dfrac{p^{-n}}{\left(p^{-\alpha\left(k+1\right)}-p^{-\alpha n}+p^{-\alpha k}\Delta_{k}\left(\alpha\right)\right)^{2}}
\]

\[
<\stackrel[n=0]{k}{\sum}\dfrac{p^{\left(2\alpha-1\right)n}}{\left(1-p^{-\alpha\left(k-n\right)}p^{-\alpha}-p^{-\alpha\left(k-n\right)}\Delta_{k}\left(\alpha\right)\right)^{2}}
\]
\[
+\dfrac{p^{\left(2\alpha-1\right)k}}{p\Delta_{k}^{2}\left(\alpha\right)}+\stackrel[n=k+2]{\infty}{\sum}\dfrac{p^{-n}}{\left(p^{-\alpha k}\Delta_{k}\left(\alpha\right)\right)^{2}}
\]

\[
<\stackrel[n=0]{k}{\sum}\dfrac{p^{\left(2\alpha-1\right)n}}{\left(1-p^{-\alpha}-\Delta_{k}\left(\alpha\right)\right)^{2}}+\dfrac{p^{\left(2\alpha-1\right)k}}{p\Delta_{k}^{2}\left(\alpha\right)}+\dfrac{p^{\left(2\alpha-1\right)k}}{p\left(p-1\right)\Delta_{k}^{2}\left(\alpha\right)}<
\]

\[
<\dfrac{p^{\left(2\alpha-1\right)k}}{\left(1-p^{-\alpha}-\Delta_{k}\left(\alpha\right)\right)^{2}}\dfrac{p^{-\left(2\alpha-1\right)k}-p^{2\alpha-1}}{1-p^{2\alpha-1}}+\dfrac{p^{\left(2\alpha-1\right)k}}{\left(p-1\right)\Delta_{k}^{2}\left(\alpha\right)}.
\]
Next we separately consider the cases of $\alpha>1$, $0<\alpha<1$,
and $\alpha=1$. For $\alpha>1$, taking account of Lemma 1, we have
\[
\stackrel[n=0]{\infty}{\sum}\dfrac{p^{-n}}{\left(p^{-\alpha n}-\mu_{k}\right)^{2}}<\dfrac{p^{\left(2\alpha-1\right)k}}{\Delta_{k}^{2}\left(\alpha\right)}\left(\dfrac{1}{p-1}+\dfrac{p^{2\alpha-1}}{p^{2\alpha-1}-1}\dfrac{\Delta_{k}^{2}\left(\alpha\right)}{\left(1-p^{-\alpha}-\Delta_{k}\left(\alpha\right)\right)^{2}}\right)
\]
\[
<p^{\left(2\alpha-1\right)k}\dfrac{\left(p+1\right)^{2}}{\left(1-p^{-\alpha}\right)^{2}\left(1-p^{-\left(\alpha-1\right)}\right)^{2}}\left(\dfrac{1}{p-1}+\left(\dfrac{p+1}{p-1}\right)^{2}\dfrac{p^{2\alpha-1}}{p^{2\alpha-1}-1}\dfrac{1}{\left(\left(1-p^{-\alpha}\right)\left(p-1\right)-1\right)^{2}}\right).
\]
For $0<\alpha<1$, taking into account Lemma 1, we have
\[
\stackrel[n=0]{\infty}{\sum}\dfrac{p^{-n}}{\left(p^{-\alpha n}-\mu_{k}\right)^{2}}<\dfrac{p^{-\left(1-2\alpha\right)k}}{\Delta_{k}^{2}\left(\alpha\right)}\left(\dfrac{1}{p-1}+\dfrac{\left(1-p^{-\alpha}\right)^{2}}{\left(1-p^{-\alpha}-\Delta_{k}\left(\alpha\right)\right)^{2}}\right)
\]
\[
<p^{k}\dfrac{\left(p+1\right)^{2}}{\left(1-p^{-\alpha}\right)^{2}\left(1-p^{-\left(1-\alpha\right)}\right)^{2}}\left(\dfrac{1}{p-1}+\left(\dfrac{\left(p-1\right)\left(1-p^{-\alpha}\right)}{\left(p-1\right)\left(1-p^{-\alpha}\right)-1}\right)^{2}\right).
\]
For $\alpha=1$, we apply Lemma 1 to obtain

\[
\stackrel[n=0]{\infty}{\sum}\dfrac{p^{-n}}{\left(p^{-n}-\mu_{k}\right)^{2}}<\stackrel[n=0]{k}{\sum}\dfrac{p^{n}}{\left(1-p^{-1}-\Delta_{k}\left(1\right)\right)^{2}}+\dfrac{p^{k}}{p\Delta_{k}^{2}\left(1\right)}+\dfrac{p^{k}}{p\left(p-1\right)\Delta_{k}^{2}\left(1\right)}
\]

\[
<\left(k+1\right)^{2}p^{k}\left(\dfrac{p+1}{p-1}\right)^{2}p^{2}\left(\dfrac{1}{p-1}+\left(\dfrac{p}{\left(p-1\right)^{2}-p}\right)^{2}\right).
\]

Let us get a lower estimate. We write

\[
\stackrel[n=0]{\infty}{\sum}\dfrac{p^{-n}}{\left(p^{-\alpha n}-\mu_{k}\right)^{2}}=\stackrel[n=0]{k}{\sum}\dfrac{p^{-n}}{\left(p^{-\alpha n}-p^{-\alpha\left(k+1\right)}-p^{-\alpha k}\Delta_{k}\left(\alpha\right)\right)^{2}}
\]
\[
+\dfrac{p^{-\left(k+1\right)}}{p^{-2\alpha k}\Delta_{k}^{2}\left(\alpha\right)}+\stackrel[n=k+2]{\infty}{\sum}\dfrac{p^{-n}}{\left(p^{-\alpha\left(k+1\right)}-p^{-\alpha n}+p^{-\alpha k}\Delta_{k}\left(\alpha\right)\right)^{2}}
\]

\[
>\dfrac{p^{\left(2\alpha-1\right)k}}{\Delta_{k}^{2}\left(\alpha\right)}\dfrac{1}{p}.
\]
Then, taking into account Lemma 1, we have
\[
\stackrel[n=0]{\infty}{\sum}\dfrac{p^{-n}}{\left(p^{-\alpha n}-\mu_{k}\right)^{2}}>p^{\left(2\alpha-1\right)k}\dfrac{\left(p+1\right)^{2}}{\left(1-p^{-\alpha}\right)^{2}\left(1-p^{-\left(\alpha-1\right)}\right)}\;\alpha>1,
\]
\[
\stackrel[n=0]{\infty}{\sum}\dfrac{p^{-n}}{\left(p^{-\alpha n}-\mu_{k}\right)^{2}}>p^{k}\dfrac{\left(p-1\right)^{2}}{p},\;0<\alpha<1,
\]
\[
\stackrel[n=0]{\infty}{\sum}\dfrac{p^{-n}}{\left(p^{-\alpha n}-\mu_{k}\right)^{2}}>\left(k+1\right)^{2}p^{k}\left(p-1\right)^{2},\;\alpha=1.
\]
Lemma 2 is proved.

\textbf{Lemma 3.} For $\alpha>1,$ there exists a finite nonzero
limit
$\underset{k\rightarrow\infty}{\lim}\Delta_{k}\left(\alpha\right)$=$\Delta\left(\alpha\right)$.

\textbf{Proof Lemma 3.} Suppose that the limit
$\underset{k\rightarrow\infty}{\lim}\Delta_{k}\left(\alpha\right)$=$\Delta\left(\alpha\right)$
does not exit. Then it follows from the Cauchy criterion that
$\exists\,\varepsilon>0\vcentcolon\forall\,K\in\mathbb{N},\exists\,k,n\in\mathbb{N},\vcentcolon\left|\Delta_{k+n}\left(\alpha\right)-\Delta_{k}\left(\alpha\right)\right|>\varepsilon.$
To prove the lemma, it is convenient to consider the sequence
$\dfrac{p^{-1}}{\Delta_{k}\left(\alpha\right)}$. Then
$\left|\dfrac{p^{-1}}{\Delta_{k}\left(\alpha\right)}-\dfrac{p^{-1}}{\Delta_{k+n}\left(\alpha\right)}\right|$
$=p^{-1}\left|\dfrac{\Delta_{k+n}\left(\alpha\right)-\Delta_{k}\left(\alpha\right)}{\Delta_{k}\left(\alpha\right)\Delta_{k+n}\left(\alpha\right)}\right|>\dfrac{p^{-1}}{\left(p-1\right)^{2}}\varepsilon.$
We apply formula (\ref{eq_delta}):

\[
\dfrac{p^{-1}}{\Delta_{k}\left(\alpha\right)}-\dfrac{p^{-1}}{\Delta_{k+n}\left(\alpha\right)}=\stackrel[i=0]{k}{\sum}\dfrac{p^{i}}{p^{\alpha i}-p^{-\alpha}-\Delta_{k}\left(\alpha\right)}-\stackrel[j=2]{\infty}{\sum}\dfrac{p^{-j}}{p^{-\alpha}-p^{-\alpha j}+\Delta_{k}\left(\alpha\right)}
\]

\[
-\stackrel[i=0]{k+n}{\sum}\dfrac{p^{i}}{p^{\alpha i}-p^{-\alpha}-\Delta_{k+n}\left(\alpha\right)}+\stackrel[j=2]{\infty}{\sum}\dfrac{p^{-j}}{p^{-\alpha}-p^{-\alpha j}+\Delta_{k+n}\left(\alpha\right)}
\]
\[
=\stackrel[i=0]{k}{\sum}p^{i}\dfrac{\Delta_{k}\left(\alpha\right)-\Delta_{k+n}\left(\alpha\right)}{\left(p^{\alpha i}-p^{-\alpha}-\Delta_{k}\left(\alpha\right)\right)\left(p^{\alpha i}-p^{-\alpha}-\Delta_{k+n}\left(\alpha\right)\right)}
\]

\[
+\stackrel[j=2]{\infty}{\sum}p^{-j}\dfrac{\Delta_{k}\left(\alpha\right)-\Delta_{k+n}\left(\alpha\right)}{\left(p^{-\alpha}-p^{-\alpha j}+\Delta_{k+n}\left(\alpha\right)\right)\left(p^{-\alpha}-p^{-\alpha j}+\Delta_{k}\left(\alpha\right)\right)}
\]

\begin{equation}
+\stackrel[i=1]{n}{\sum}\dfrac{p^{k}p^{i}}{p^{\alpha k}p^{\alpha i}-p^{-\alpha}-\Delta_{k+n}\left(\alpha\right)}.\label{expr_add}
\end{equation}
The expression in the last row in (\ref{expr_add}) can be made
arbitrarily small in view of the boundedness of
$\Delta_{k}\left(\alpha\right)$ for sufficiently large $K$. The
expression in the penultimate row in (\ref{expr_add}) can be
represented as
\[
\left[\stackrel[i=0]{k}{\sum}p^{i}\dfrac{1}{\left(p^{\alpha i}-p^{-\alpha}-\Delta_{k}\left(\alpha\right)\right)\left(p^{\alpha i}-p^{-\alpha}-\Delta_{k+n}\left(\alpha\right)\right)}\right.
\]

\begin{equation}
\left.+\stackrel[j=2]{\infty}{\sum}p^{-j}\dfrac{1}{\left(p^{-\alpha}-p^{-\alpha j}+\Delta_{k+n}\left(\alpha\right)\right)\left(p^{-\alpha}-p^{-\alpha j}+\Delta_{k}\left(\alpha\right)\right)}\right]\left(\Delta_{k}\left(\alpha\right)-\Delta_{k+n}\left(\alpha\right)\right).\label{expr_add_1}
\end{equation}
Since the expression in square brackets is positive and
$\dfrac{p^{-1}}{\Delta_{k}}-\dfrac{p^{-1}}{\Delta_{k+n}\left(\alpha\right)}$
and
$\Delta_{k}\left(\alpha\right)-\Delta_{k+n}\left(\alpha\right)$
are opposite in sign, there always exist $k,n\in\mathbb{N}$ and
$k>K$, which leads to a contradiction due to the fact that
equality (\ref{expr_add}) is violated; thus, Lemma 3 is proved.

\textbf{Lemma 4.} For $\alpha=1$ there exists a finite limit
$\underset{k\rightarrow\infty}{\lim}k\Delta_{k}\left(\alpha\right)>0$.

\textbf{Proof Lemma 4. } Suppose that the limit
$\underset{k\rightarrow\infty}{\lim}k\Delta_{k}\left(1\right)$
does not exist. Then it follows from the Cauchy criterion that
$\exists\,\varepsilon>0\vcentcolon\:\forall\,K\in\mathbb{N},\exists\,k,n\in\mathbb{N}\vcentcolon$
$\left|\left(k+n\right)\Delta_{k+n}\left(1\right)-k\Delta_{k}\left(1\right)\right|>\varepsilon$.
To prove the lemma, it is convenient to consider the sequence
$\dfrac{p^{-1}}{k\Delta_{k}\left(1\right)}$. Then
\[
\left|\dfrac{p^{-1}}{k\Delta_{k}\left(1\right)}-\dfrac{p^{-1}}{\left(k+n\right)\Delta_{k+n}\left(1\right)}\right|=p^{-1}\left|\dfrac{\left(k+n\right)\Delta_{k+n}\left(1\right)-k\Delta_{k}\left(1\right)}{k\Delta_{k}\left(1\right)\left(k+n\right)\Delta_{k+n}\left(1\right)}\right|>\dfrac{p^{-1}}{\left(p-1\right)^{2}}\varepsilon.
\]
 We apply formula (\ref{eq_delta}):

\[
\dfrac{p^{-1}}{k\Delta_{k}\left(1\right)}-\dfrac{p^{-1}}{\left(k+n\right)\Delta_{k+n}\left(1\right)}=\stackrel[i=0]{k}{\sum}\dfrac{p^{i}}{k\left(p^{i}-p^{-1}-\Delta_{k}\left(1\right)\right)}-\stackrel[j=2]{\infty}{\sum}\dfrac{p^{-j}}{k\left(p^{-1}-p^{-j}+\Delta_{k}\left(1\right)\right)}
\]

\[
-\stackrel[i=0]{k+n}{\sum}\dfrac{p^{i}}{\left(k+n\right)\left(p^{i}-p^{-1}-\Delta_{k+n}\left(1\right)\right)}+\stackrel[j=2]{\infty}{\sum}\dfrac{p^{-j}}{\left(k+n\right)\left(p^{-1}-p^{-j}+\Delta_{k+n}\left(1\right)\right)}
\]

\[
=\stackrel[i=0]{k}{\sum}p^{i}\dfrac{k\Delta_{k}\left(1\right)-\left(k+n\right)\Delta_{k+n}\left(\alpha\right)}{k\left(k+n\right)\left(p^{i}-p^{-1}-\Delta_{k}\left(1\right)\right)\left(p^{i}-p^{-1}-\Delta_{k+n}\left(1\right)\right)}
\]
\[
+\stackrel[j=2]{\infty}{\sum}p^{-j}\dfrac{k\Delta_{k}\left(\alpha\right)-\left(k+n\right)\Delta_{k+n}\left(\alpha\right)}{k\left(k+n\right)\left(p^{-1}-p^{-j}+\Delta_{k+n}\left(1\right)\right)\left(p^{-1}-p^{-j}+\Delta_{k}\left(1\right)\right)}
\]
\[
+\dfrac{1}{k+n}\stackrel[i=1]{n}{\sum}\dfrac{p^{k}p^{i}}{\left(p^{k}p^{i}-p^{-1}-\Delta_{k+n}\left(1\right)\right)}
\]

\[
+\dfrac{n}{k\left(k+n\right)}\left[\stackrel[i=0]{k}{\sum}\dfrac{p^{i}\left(p^{i}-p^{-1}\right)}{\left(p^{i}-p^{-1}-\Delta_{k+n}\left(1\right)\right)\left(p^{i}-p^{-1}-\Delta_{k}\left(1\right)\right)}\right.
\]

\begin{equation}
\left.-\stackrel[j=2]{\infty}{\sum}\dfrac{p^{-j}\left(p^{-1}-p^{-j}\right)}{\left(p^{-1}-p^{-j}+\Delta_{k}\left(1\right)\right)\left(p^{-1}-p^{-j}+\Delta_{k}\left(1\right)\right)}\right].\label{expr_add_2}
\end{equation}
For sufficiently large $K$ the terms in the last two rows
(\ref{expr_add_2}) can be made arbitrarily small. Indeed, taking
into account that
$\underset{k\rightarrow\infty}{\lim}\Delta_{k}\left(1\right)=0$,
we have

\[
\dfrac{1}{k+n}\stackrel[i=1]{n}{\sum}\dfrac{p^{k}p^{i}}{\left(p^{k}p^{i}-p^{-1}-\Delta_{k+n}\left(1\right)\right)}\sim\dfrac{1}{k},
\]

\[
\dfrac{n}{k\left(k+n\right)}\stackrel[j=2]{\infty}{\sum}\dfrac{p^{-j}\left(p^{-1}-p^{-j}\right)}{\left(p^{-1}-p^{-j}+\Delta_{k}\left(1\right)\right)\left(p^{-1}-p^{-j}+\Delta_{k}\left(1\right)\right)}\sim\dfrac{1}{k^{2}},
\]

\[
\dfrac{n}{k\left(k+n\right)}\stackrel[i=0]{k}{\sum}\dfrac{p^{i}\left(p^{i}-p^{-1}\right)}{\left(p^{i}-p^{-1}-\Delta_{k}\left(1\right)\right)\left(p^{i}-p^{-1}-\Delta_{k+1}\left(1\right)\right)}
\]

\[
=\dfrac{n}{k\left(k+n\right)}\left[\stackrel[i=0]{k}{\sum}\left(\dfrac{p^{i}}{\left(p^{i}-p^{-1}-\Delta_{k}\left(1\right)\right)}\right)+\Delta_{k}\left(1\right)\stackrel[i=0]{k}{\sum}\dfrac{p^{i}}{\left(p^{i}-p^{-1}-\Delta_{k}\left(1\right)\right)\left(p^{i}-p^{-1}-\Delta_{k}\left(1\right)\right)}\right]
\]

\[
=\dfrac{n}{k\left(k+n\right)}\left[\stackrel[i=0]{k}{\sum}\left(1+\dfrac{p^{-1}+\Delta_{k}}{\left(p^{i}-p^{-1}-\Delta_{k}\left(1\right)\right)}\right)+\Delta_{k}\left(1\right)\stackrel[i=0]{k}{\sum}\dfrac{p^{i}}{\left(p^{i}-p^{-1}-\Delta_{k}\left(1\right)\right)\left(p^{i}-p^{-1}-\Delta_{k}\left(1\right)\right)}\right]
\]

\[
\sim\dfrac{1}{k}.
\]
Then

\[
\dfrac{p^{-1}}{k\Delta_{k}\left(1\right)}-\dfrac{p^{-1}}{\left(k+n\right)\Delta_{k+n}\left(1\right)}=\dfrac{k\Delta_{k}\left(1\right)-\left(k+n\right)\Delta_{k}\left(1\right)}{k\left(k+n\right)}
\]
\[
\times\left[\stackrel[i=0]{k}{\sum}\dfrac{p^{i}}{\left(p^{i}-p^{-1}-\Delta_{k}\left(1\right)\right)\left(p^{i}-p^{-1}-\Delta_{k+n}\left(1\right)\right)}+\stackrel[j=2]{\infty}{\sum}\dfrac{p^{-j}}{\left(p^{-1}-p^{-j}+\Delta_{k+n}\left(1\right)\right)\left(p^{-1}-p^{-j}+\Delta_{k}\left(1\right)\right)}\right]
\]

\begin{equation}
+o\left(1\right).\label{expr_add_3}
\end{equation}
Since the expression in square brackets in (\ref{expr_add_3}) is
positive and the expressions
$\dfrac{p^{-1}}{k\Delta_{k}}-\dfrac{p^{-1}}{\left(k+n\right)\Delta_{k+n}\left(\alpha\right)}$
and
$k\Delta_{k}\left(\alpha\right)-\left(k+n\right)\Delta_{k+n}\left(\alpha\right)$
are of opposite sign, there always exist $k,n\in\mathbb{N}$ and
$k>K$ such that the equality (\ref{expr_add_3}) is contradictory.
This proves Lemma 4.

\textbf{Lemma 5.} For $0<\alpha<1$ there exists a finite limit
$\underset{k\rightarrow\infty}{\lim}p^{\left(1-\alpha\right)k}\Delta_{k}\left(\alpha\right)>0$.

\textbf{Proof Lemma 5.} By estimate (\ref{Eq_lemma_1}) in Lemma 1,

\[
\dfrac{1}{p+1}\dfrac{p^{\alpha-1}-1}{p^{\alpha-1}}\dfrac{p^{\alpha}-1}{p^{\alpha}}<p^{\left(1-\alpha\right)k}\Delta_{k}\left(\alpha\right)<\dfrac{1}{p-1},\;0<\alpha<1,
\]
the sequence
$p^{\left(1-\alpha\right)k}\Delta_{k}\left(\alpha\right)$ is
bounded. Then, to prove the existence of the limit, it suffices to
prove the monotonicity of this sequence. From equation
$J_{r}\left(-\lambda_{k}\right)=0$ we have

\[
\stackrel[n=0]{\infty}{\sum}\dfrac{p^{-n}}{p^{-\alpha n}-p^{-\alpha\left(k+1\right)}-p^{-\alpha k}\Delta_{k}\left(\alpha\right)}=0,
\]
whence
\[
\dfrac{p^{-1}}{p^{\left(1-\alpha\right)k}\Delta_{k}}=\stackrel[j=0]{k}{\sum}\dfrac{p^{-j}p^{k}}{p^{-\alpha j}p^{k}-p^{\left(1-\alpha\right)k}p^{-\alpha}-p^{\left(1-\alpha\right)k}\Delta_{k}}
\]

\[
-\stackrel[j=2]{\infty}{\sum}\dfrac{p^{-j}}{p^{\left(1-\alpha\right)k}p^{-\alpha}+p^{\left(1-\alpha\right)k}\Delta_{k}-p^{-\alpha j}p^{\left(1-\alpha\right)k}}.
\]
 Consider the difference
\[
\dfrac{p^{-1}}{p^{\left(1-\alpha\right)k}\Delta_{k}}-\dfrac{p^{-1}}{p^{\left(1-\alpha\right)\left(k+n\right)}\Delta_{k+n}}=-\dfrac{1}{p^{\left(1-\alpha\right)k}}\stackrel[i=1]{n}{\sum}\dfrac{p^{-i}}{p^{-\alpha i}-p^{-\alpha}-\Delta_{k}}
\]

\[
+\stackrel[i=0]{k}{\sum}\left(\dfrac{p^{-i}}{p^{-\alpha i}-p^{-\alpha k}\left(p^{-\alpha}+\Delta_{k}\right)}-\dfrac{p^{-i}}{p^{-\alpha i}-p^{-\alpha\left(k+n\right)}\left(p^{-\alpha}+\Delta_{k+n}\right)}\right)
\]
\begin{equation}
-\dfrac{1}{p^{\left(1-\alpha\right)k}}\stackrel[j=2]{\infty}{\sum}\left(\dfrac{p^{-j}}{p^{-\alpha}-p^{-\alpha
j}+\Delta_{k}}-\dfrac{p^{-j}}{p^{\left(1-\alpha\right)n}\left(p^{-\alpha}-p^{-\alpha
j}+\Delta_{k+n}\right)}\right).\label{eq_1}
\end{equation}
For sufficiently  large $k>K$ the first and third terms in
(\ref{eq_1}) can be made arbitrarily small due to the factor
$p^{\left(1-\alpha\right)k}$ in the denominator. Next, we assume
that there exists at least one $k$, $k>K$, such that the equality
\begin{equation}
p^{\left(1-\alpha\right)\left(k+n\right)}\Delta_{k+n}<p^{\left(1-\alpha\right)k}\Delta_{k}\label{pred}
\end{equation}
holds. Then
$\dfrac{p^{-1}}{p^{\left(1-\alpha\right)k}\Delta_{k}}-\dfrac{p^{-1}}{p^{\left(1-\alpha\right)\left(k+n\right)}\Delta_{k+n}}<0$.
In this case the second term is positive because the denominators
satisfy the following inequality due to (\ref{pred}):
\[
p^{-\alpha i}-p^{-\alpha
k}\left(p^{-\alpha}+\Delta_{k}\right)>p^{-\alpha
i}-p^{-\alpha\left(k+n\right)}\left(p^{-\alpha}+\Delta_{k+n}\right),
\]
which leads to
$\dfrac{p^{-1}}{p^{\left(1-\alpha\right)k}\Delta_{k}}-\dfrac{p^{-1}}{p^{\left(1-\alpha\right)\left(k+n\right)}\Delta_{k+n}}\geq0$.
The contradiction obtained shows that there does not exist a $k>K$
such that (\ref{pred}) holds. As a result,
$p^{\left(1-\alpha\right)k}\Delta_{k}$ is a nondecreasing sequence
for large enough $k$, and hence it is monotonic. Since it is
bounded by nonnegative numbers, it has a finite limit. Lemma 5 is
proved.

Let us return to the proof of Theorem 9. It is easily seen that

\begin{equation}
\underset{k\rightarrow\infty}{\lim}\lambda_{k}p^{\alpha k}=\left\{ \begin{array}{c}
p^{-\alpha r}\left(p^{-\alpha}+\Delta\left(\alpha\right)\right),\;\alpha>1,\\
p^{-\alpha r}p^{-\alpha},\;\alpha=1,\\
p^{-\alpha r}p^{-\alpha},\;0<\alpha<1,
\end{array}\right.\label{lim_lamda}
\end{equation}

\begin{equation}
\underset{k\rightarrow\infty}{\lim}b_{k}\dfrac{p^{\left(2\alpha-1\right)k}}{\Delta_{k}^{2}\left(\alpha\right)}=\left\{ \begin{array}{c}
p^{-\alpha r}I_{0}^{\nu-r-1}\left(0\right)\dfrac{1}{\Delta^{2}\left(\alpha\right)\stackrel[n=-\infty]{\infty}{\sum}\dfrac{p^{n}}{\left(p^{\alpha n}-p^{-\alpha}-\Delta\left(\alpha\right)\right)^{2}}},,\;\alpha>1,\\
p^{-\alpha r}I_{0}^{\nu-r-1}\left(0\right)p^{-1},\;\alpha=1,\\
p^{-\alpha r}I_{0}^{\nu-r-1}\left(0\right)p^{-1},\;0<\alpha<1,
\end{array}\right.\label{est_app}
\end{equation}
where $\Delta\left(\alpha\right)$ satisfies the estimate

\[
\dfrac{\left(1-p^{-\alpha}\right)\left(1-p^{-\left(\alpha-1\right)}\right)}{p+1}<\Delta\left(\alpha\right)<\dfrac{1}{p-1}.
\]
The calculation of the limits (\ref{lim_lamda}) is trivial. Let us
show the scheme for calculating the limits (\ref{est_app}). We
have

\[
\underset{k\rightarrow\infty}{\lim}b_{k}\dfrac{p^{\left(2\alpha-1\right)k}}{\Delta_{k}^{2}\left(\alpha\right)}=p^{-\alpha r}I_{0}^{\nu-r-1}\left(0\right)\underset{k\rightarrow\infty}{\lim}\dfrac{p^{\left(2\alpha-1\right)k}}{\Delta_{k}^{2}\left(\alpha\right)\stackrel[k=0]{\infty}{\sum}\dfrac{p^{-n}}{\left(p^{-\alpha n}-p^{-\alpha k}\left(p^{-\alpha}+\Delta_{k}\left(\alpha\right)\right)\right)^{2}}}.
\]
Consider the limit of the denominator. For $0<\alpha\leq1$ there
holds
$\underset{k\rightarrow\infty}{\lim}\Delta_{k}\left(\alpha\right)=0$;
therefore,
\[
\underset{k\rightarrow\infty}{\lim}p^{-\left(2\alpha-1\right)k}\Delta_{k}^{2}\left(\alpha\right)\left(\stackrel[n=0]{k}{\sum}\dfrac{p^{-n}}{\left(p^{-\alpha n}-p^{-\alpha\left(k+1\right)}-p^{-\alpha k}\Delta_{k}\left(\alpha\right)\right)^{2}}+\dfrac{p^{-\left(k+1\right)}}{p^{-2\alpha k}\Delta_{k}^{2}\left(\alpha\right)}\right.
\]
\[
\left.+\dfrac{p^{-\left(k+1\right)}}{p^{-2\alpha k}\Delta_{k}^{2}\left(\alpha\right)}+\stackrel[n=k+2]{\infty}{\sum}\dfrac{p^{-n}}{\left(p^{-\alpha\left(k+1\right)}-p^{-\alpha n}+p^{-\alpha k}\Delta_{k}\left(\alpha\right)\right)^{2}}\right)
\]

\[
=p^{-1}+\underset{k\rightarrow\infty}{\lim}p^{-\left(2\alpha-1\right)k}\Delta_{k}^{2}\left(\alpha\right)\left(\stackrel[n=0]{k}{\sum}\dfrac{p^{-n}}{\left(p^{-\alpha n}-p^{-\alpha\left(k+1\right)}-p^{-\alpha k}\Delta_{k}\left(\alpha\right)\right)^{2}}\right.
\]
\[
\left.+\stackrel[n=k+2]{\infty}{\sum}\dfrac{p^{-n}}{\left(p^{-\alpha\left(k+1\right)}-p^{-\alpha n}+p^{-\alpha k}\Delta_{k}\left(\alpha\right)\right)^{2}}\right)
\]

\[
=p^{-1}+\underset{k\rightarrow\infty}{\lim}\Delta_{k}^{2}\left(\alpha\right)\left(\stackrel[i=0]{k}{\sum}\dfrac{p^{i}}{\left(p^{\alpha
i}-p^{-\alpha}-\Delta_{k}\left(\alpha\right)\right)^{2}}+\stackrel[j=k+2]{\infty}{\sum}\dfrac{p^{-j}}{\left(p^{-\alpha
j}-p^{-\alpha}+\Delta_{k}\left(\alpha\right)\right)^{2}}\right)=p^{-1}.
\]
If $\alpha>1$, then
$\underset{k\rightarrow\infty}{\lim}\Delta_{k}\left(\alpha\right)=\Delta\left(\alpha\right)>0$,
and the limit of the denominator is

\[
p^{-1}+\underset{k\rightarrow\infty}{\lim}\Delta_{k}^{2}\left(\alpha\right)\left(\stackrel[i=0]{k}{\sum}\dfrac{p^{i}}{\left(p^{\alpha i}-p^{-\alpha}-\Delta_{k}\left(\alpha\right)\right)^{2}}+\stackrel[j=k+2]{\infty}{\sum}\dfrac{p^{-j}}{\left(p^{-\alpha j}-p^{-\alpha}+\Delta_{k}\left(\alpha\right)\right)^{2}}\right)
\]

\[
=p^{-1}+\Delta^{2}\left(\alpha\right)\left(\stackrel[i=0]{k}{\sum}\dfrac{p^{i}}{\left(p^{\alpha i}-p^{-\alpha}-\Delta\left(\alpha\right)\right)^{2}}+\stackrel[j=k+2]{\infty}{\sum}\dfrac{p^{i}}{\left(p^{-\alpha i}-p^{-\alpha}+\Delta\left(\alpha\right)\right)^{2}}\right)
\]

\[
=p^{-1}+\Delta^{2}\left(\alpha\right)\left(\stackrel[i=0]{\infty}{\sum}\dfrac{p^{i}}{\left(p^{\alpha i}-p^{-\alpha}-\Delta\left(\alpha\right)\right)^{2}}\right.
\]
\[
\left.+\stackrel[i=-\infty]{-1}{\sum}\dfrac{p^{i}}{\left(p^{\alpha i}-p^{-\alpha}+\Delta\left(\alpha\right)\right)^{2}}-\dfrac{p^{-1}}{\left(p^{-\alpha}-p^{-\alpha}-\Delta\left(\alpha\right)\right)^{2}}\right)
\]

\[
=\Delta^{2}\left(\alpha\right)\stackrel[i=-\infty]{\infty}{\sum}\dfrac{p^{i}}{\left(p^{\alpha i}-p^{-\alpha}-\Delta\left(\alpha\right)\right)^{2}}.
\]

To complete the proof of Theorem 9, we need Lemma 6.

\textbf{Lemma 6. } Let $a_{n}$, $b_{n}$, and $c_{n}$,
$n=1,2,\ldots$, be three infinite positive sequences in
$\mathbb{R}$, and let there exist numbers  $a>1$, $b>1$, and
$k\in\mathbb{Z}_{+}$ such that
\begin{equation}
a_{n}a^{n}\sim1,\;b_{n}b^{n}\sim1,\;c_{n}n^{k}\sim1\label{a_n_b_n}
\end{equation}
as $n\rightarrow\infty$. Then the series

\begin{equation}
S\left(t\right)=\mathop{\sum}\limits _{n=1}^{\infty}a_{n}c_{n}e^{-b_{n}t}\label{Ser_S}
\end{equation}
exhibits the following asymptotic behavior as
$t\rightarrow\infty$:
\[
S\left(t\right)\sim\left(\log t\right)^{-k}t^{-\tfrac{\log a}{\log b}}f\left(t\right),
\]
where the function $f\left(t\right)$ is log-periodic,
$f\left(bt\right)=f\left(t\right)$, and has the form
\begin{equation}
f\left(t\right)=\left(\log b\right)^{k-1}\sum_{m=-\infty}^{\infty}\exp\left(\dfrac{2\pi im}{\log b}\log t\right)\Gamma\left(\dfrac{\log a}{\log b}-\dfrac{2\pi im}{\log b}\right).\label{S_th}
\end{equation}

\textbf{Proof Lemma 6. } Lemma 6 for $k=0$ was proved as a theorem
in \cite{BZ_2024} and can easily be generalized to the case of
$k>1$; therefore, here we restrict ourselves to the scheme of the
proof.

Take functions $a\left(x\right)$, $b\left(x\right)$, and
$c\left(x\right)$ from the class
$C^{\infty}\left(\mathbb{R}\right)$ such that the following
conditions are satisfied:
\[
a\left(n\right)=a_{n},\;b\left(n\right)=b_{n},\;c\left(n\right)=c_{n}\;\mathrm{for}\;n\geq1,
\]
\[
a\left(x\right)=a^{-x},\;b\left(x\right)=b^{-x},\;c\left(x\right)=1\;\mathrm{for}\;x<1,
\]
\[
\lim_{x\rightarrow+\infty}a\left(x\right)a^{x}=\lim_{x\rightarrow+\infty}b\left(x\right)b^{x}=\lim_{x\rightarrow+\infty}c\left(x\right)x^{k}=1.
\]
Then
\begin{equation}
a\left(x\right)=a^{-x}\left(1+\varepsilon\left(x\right)\right),\label{app_a(x)}
\end{equation}
\begin{equation}
b\left(x\right)=b^{-x}\left(1+\delta\left(x\right)\right),\label{app_b(x)}
\end{equation}
\begin{equation}
c\left(x\right)=x^{-k}\left(1+\epsilon\left(x\right)\right)\Theta\left(x-1\right)+\Theta\left(1-x\right),\label{app_c(x)}
\end{equation}
where $\Theta\left(x\right)=\left\{ \begin{array}{c}
1,\:x\geq0,\\
0,\:x<0,
\end{array}\right.$ and the functions $\varepsilon\left(x\right)$, $\delta\left(x\right)$, and
$\epsilon\left(x\right)$ are infinitesimal as
$x\rightarrow+\infty$.

Consider the series
\begin{equation}
Q\left(t\right)=\mathop{\sum}\limits _{m=-\infty}^{\infty}a\left(m\right)c\left(m\right)e^{-b\left(m\right)t}=S\left(t\right)+R\left(t\right),\label{app_Q}
\end{equation}
where
\[
R\left(t\right)=\mathop{\sum}\limits _{i=0}^{\infty}a^{i}e^{-b^{i}t}.
\]
Following \cite{BZ_2024}, we can show that

\begin{equation}
Q\left(t\right)=\sum_{m=-\infty}^{\infty}\intop_{-\infty}^{\infty}a\left(x\right)c\left(x\right)e^{-b\left(x\right)t}\exp\left(-2\pi imx\right)dx.\label{app_Q(t)}
\end{equation}
Changing the variable $x\rightarrow y=-x\log b+\log t$ in
(\ref{app_Q(t)}) and using (\ref{app_a(x)})--(\ref{app_c(x)}), we
write

\begin{equation}
Q\left(t\right)=\dfrac{1}{\log b}t^{-\tfrac{\log a}{\log b}}\mathop{\sum}\limits _{k=-\infty}^{+\infty}t^{\tfrac{2\pi ik}{\log b}}Q_{k}\left(t\right),\label{app_Q_t}
\end{equation}

\noindent where
\[
Q_{k}\left(t\right)=\mathop{\smallint}\limits _{-\infty}^{\infty}dy\exp\left(-\dfrac{2\pi ik}{\log b}y\right)
\]
\[
\times\left(\left(\dfrac{\log t-y}{\log b}\right)^{-k}\left(1+\epsilon\left(\dfrac{\log t-y}{\log b}\right)\right)\Theta\left(\dfrac{\log t-y}{\log b}-1\right)+\Theta\left(1-\dfrac{\log t-y}{\log b}\right)\right)
\]

\noindent
\begin{equation}
\times\left(1+\varepsilon\left(-\dfrac{y}{\log b}+\dfrac{\log t}{\log b}\right)\right)\exp\left(-e^{y}\delta\left(-\dfrac{y}{\log b}+\dfrac{\log t}{\log b}\right)\right)\exp\left(\dfrac{\log a}{\log b}y-e^{y}\right).\label{app_Q_k_y}
\end{equation}
The integral (\ref{app_Q_k_y}) converges absolutely for any
$t>T\in\mathbb{R}_{+}$ and is the Fourier integral of a function
from the Schwartz space. Therefore, $Q_{k}\left(t\right)$ for any
$t>T\in\mathbb{R}_{+}$ is also a function of $k$ from the Schwartz
space. This means that the series
\begin{equation}
\log bt^{\tfrac{\log a}{\log b}}Q\left(t\right)=\mathop{\sum}\limits _{k=-\infty}^{+\infty}t^{\tfrac{2\pi ik}{\log b}}Q_{k}\left(t\right)\label{app_series_unif_t}
\end{equation}
converges absolutely for any $t>T\in\mathbb{R}_{+}$; i.e., the
series
\[
\mathop{\sum}\limits _{k=-\infty}^{+\infty}\left|Q_{k}\left(t\right)\right|.
\]
converges. Since $Q_{k}\left(t\right)$ is bounded in $t$, it
follows that
\[
\mathop{\sum}\limits _{k=-\infty}^{+\infty}\left|Q_{k}\left(t\right)\right|\leq\mathop{\sum}\limits _{k=-\infty}^{+\infty}\sup_{t>T}\left|Q_{k}\left(t\right)\right|,
\]
which implies that the series $\mathop{\sum}\limits
_{k=-\infty}^{+\infty}t^{\tfrac{2\pi ik}{\log
b}}Q_{k}\left(t\right)$ is bounded and hence is uniformly
convergent in $t>T$. It is easily seen that the series
$t^{\tfrac{\log a}{\log b}}R\left(t\right)=\mathop{\sum}\limits
_{i=1}^{\infty}a^{i} \exp\left(-b^{i}t+\log t\dfrac{\log a}{\log
b}\right)$ converges uniformly in $t$ and has zero limit as
$t\rightarrow\infty$. Hence it follows that
\begin{equation}
Q\left(t\right)\sim S\left(t\right)\label{app_QS}
\end{equation}
as $t\rightarrow\infty$.

\noindent Next, consider the series

\begin{equation}
P\left(t\right)=\mathop{\sum}\limits _{m=-\infty}^{\infty}a^{-i}\left(i^{-k}\Theta\left(i-1\right)+\Theta\left(1-1\right)\right)e^{-b^{-i}t},\label{app_P}
\end{equation}
which is a particular case of the series (\ref{app_Q}). Taking
account of (\ref{app_Q_t})--(\ref{app_Q_k_y}), we write

\[
P\left(t\right)=\left(\log b\right)^{k-1}t^{-\tfrac{\log a}{\log b}}\mathop{\sum}\limits _{k=-\infty}^{+\infty}t^{\tfrac{2\pi ik}{\log b}}
\]

\noindent
\[
\times\mathop{\smallint}\limits _{-\infty}^{\infty}dy\exp\left(\dfrac{\log a}{\log b}y-e^{y}\right)\exp\left(-\dfrac{2\pi ik}{\log b}y\right)
\]

\noindent
\begin{equation}
\times\left(\left(\log t-y\right)^{-k}\Theta\left(\dfrac{\log t-y}{\log b}-1\right)+\Theta\left(1-\dfrac{\log t-y}{\log b}\right)\right),\label{app_P_k}
\end{equation}

\noindent where series (\ref{app_P_k}) converges uniformly in
$t>T\in\mathbb{R}_{+}$. Then
\begin{equation}
P\left(t\right)\sim\left(\log b\right)^{k-1}\left(\log t\right)^{-k}t^{-\tfrac{\log a}{\log b}}\mathop{\sum}\limits _{k=-\infty}^{+\infty}t^{\tfrac{2\pi ik}{\log b}}P_{k}\left(t\right)\label{app_P(t)}
\end{equation}
as $t\rightarrow\infty$, where

\noindent
\begin{equation}
P_{k}\left(t\right)=\mathop{\smallint}\limits _{-\infty}^{\infty}dy\exp\left(\dfrac{\log a}{\log b}y-e^{y}\right)\exp\left(-\dfrac{2\pi ik}{\log b}y\right)=\Gamma\left(\dfrac{\log a}{\log b}-\dfrac{2\pi im}{\log b}\right).\label{P_k}
\end{equation}

\noindent Consider the limit of the difference
\[
\lim_{t\rightarrow\infty}\left(\left(\log b\right)^{1-k}\left(\log
t\right)^{k}t^{\tfrac{\log a}{\log b}}Q\left(t\right)-\left(\log
b\right)^{1-k}\left(\log t\right)^{k}t^{\tfrac{\log a}{\log
b}}P\left(t\right)\right)=\]

\noindent
\[
=\lim_{t\rightarrow\infty}\mathop{\sum}\limits
_{k=-\infty}^{+\infty}t^{\tfrac{2\pi ik}{\log
b}}\mathop{\smallint}\limits
_{-\infty}^{\infty}dy\exp\left(-\dfrac{2\pi ik}{\log
b}y\right)\exp\left(-e^{y}-\dfrac{\log a}{\log b}y-\dfrac{2\pi
ik}{\log b}y\right)\]
\[
\times\left[\left(\dfrac{\log t}{\log
b}\right)^{k}\left(\left(\dfrac{\log t-y}{\log
b}\right)^{-k}\left(1+\epsilon\left(\dfrac{\log t-y}{\log
b}\right)\right)\Theta\left(\dfrac{\log t-y}{\log
b}-1\right)+\Theta\left(1-\dfrac{\log t-y}{\log
b}\right)\right)\right.
\]

\noindent
\begin{equation}
\left.\times\left(1+\varepsilon\left(-\dfrac{y}{\log
b}+\dfrac{\log t}{\log
b}\right)\right)\exp\left(-e^{y}\delta\left(-\dfrac{y}{\log
b}+\dfrac{\log t}{\log
b}\right)\right)-1\right].\label{app_lim_diff}
\end{equation}

\noindent In view of the uniform convergence in $t$, the limit can
be taken under the summation sign. Since, for any $y$,
\[
\lim_{t\rightarrow\infty}t^{\tfrac{2\pi ik}{\log b}}\left[\left(\dfrac{\log t}{\log b}\right)^{k}\left(\left(\dfrac{\log t-y}{\log b}\right)^{-k}\left(1+\epsilon\left(\dfrac{\log t-y}{\log b}\right)\right)\Theta\left(\dfrac{\log t-y}{\log b}-1\right)+\Theta\left(1-\dfrac{\log t-y}{\log b}\right)\right)\right.
\]

\noindent
\[
\left.\times\left(1+\varepsilon\left(-\dfrac{y}{\log
b}+\dfrac{\log t}{\log
b}\right)\right)\exp\left(-e^{y}\delta\left(-\dfrac{y}{\log
b}+\dfrac{\log t}{\log b}\right)\right)-1\right]=0,
\]

\noindent the limit (\ref{app_lim_diff}) is zero; hence,
\[
\lim_{t\rightarrow\infty}\left(\dfrac{Q\left(t\right)}{P\left(t\right)}-1\right)=\lim_{t\rightarrow\infty}\dfrac{\left(\log b\right)^{1-k}\left(\log t\right)^{k}t^{\tfrac{\log a}{\log b}}Q\left(t\right)-\left(\log b\right)^{1-k}\left(\log t\right)^{k}t^{\tfrac{\log a}{\log b}}P\left(t\right)}{\left(\log b\right)^{1-k}\left(\log t\right)^{k}t^{\tfrac{\log a}{\log b}}P\left(t\right)}=0.
\]
The last equality implies that $Q\left(t\right)\sim
P\left(t\right)$, whence, taking into account (\ref{app_QS}), we
obtain
\[
S\left(t\right)\sim P\left(t\right).
\]

\noindent This, together with (\ref{app_P(t)})--(\ref{app_P_k}),
proves the assertion of Lemma 6.

In view of Lemma 6 the assertions of (\ref{as_a<1}),
(\ref{as_a=00003D1}), and (\ref{as_a>1}) of Theorem 8 are a direct
consequence of formulas (\ref{S_th}), (\ref{f(t)_sol}),
(\ref{lim_lamda}), and (\ref{est_app}). Theorem 9 is proved.

Notice that asymptotics similar to the asymptotics (\ref{as_a<1}),
(\ref{as_a=00003D1}), and (\ref{as_a>1}) can also be obtained for
the probability density $f_{ret}\left(t\right)$ of the first
return time.

\section{Problem of the number of hittings the domain $B_{r}\left(a\right)$ by the trajectories
of a $p$-adic stochastic process.}

In the previous section we have shown that for $\alpha>1$ the
trajectory of the stochastic process $\xi\left(t,\omega\right)$
reaches the domain $B_{r}\left(a\right)$ infinitely many times. In
this case, starting from the first passage time, one can consider
the second passage as the first return, taking the first passage
time as the initial time and the uniform distribution in the
domain $B_{r}\left(a\right)$ as the initial distribution. In this
section we answer the question of how the average number of
hittings the domain $B_{r}\left(a\right)$ increases with
increasing the observation time, as well as how the distribution
function of the $m$th hitting the domain $B_{r}\left(a\right)$
varies in time.

Let
$N_{B_{r}\left(a\right)}\left(t,\omega\right)\vcentcolon\Omega\times\mathbb{R}_{+}\rightarrow\mathbb{Z}_{+}$
be a stochastic process of the number of hittings the domain
$B_{r}\left(a\right)$ by the trajectories of the stochastic
process $\xi\left(t,\omega\right)$ within the time interval
$\left(0,t\right]$. Let us calculate the probability of the $m$th
hitting the domain $B_{r}\left(a\right)$ by the trajectory of the
process $\xi\left(t,\omega\right)$. Suppose that
$Q_{t}^{m}=\left\{ \omega\in\Omega\vcentcolon
N_{B_{r}\left(a\right)}\left(t,\omega\right)\geq m\right\} $ is an
event such that, within the time interval $\left(0,t\right]$, the
trajectory of the stochastic process $\xi\left(t,\omega\right)$
hits the domain $B_{r}\left(a\right)$ at least $m$ times. It is
obvious that $Q_{0}^{m}=\varnothing$ and $Q_{t}^{0}=\Omega$.

Introduce the notation
$q^{\left(m\right)}\left(t\right)=\mathrm{P}\left\{
Q_{t}^{m}\right\} $ for the probability of the $m$th hitting the
domain $B_{r}\left(a\right)$ by the trajectory within the time
interval $\left(0,t\right]$.

\textbf{Theorem 10. } Let $f\left(t\right)$ be the probability
density of the random variable
$\tau_{B_{r}\left(a\right)}\left(\omega\right)$ -- the first
passage time to the domain $B_{r}\left(a\right)$, and let
$f_{ret}\left(t\right)$ be the probability density of the random
variable $\tau_{B_{r}\left(a\right)}^{ret}\left(\omega\right)$ --
the first return time to the domain $B_{r}\left(a\right)$. Then
the probability $q^{\left(m\right)}\left(t\right)$ satisfies the
following recurrence equation:

\begin{equation}
q^{\left(0\right)}\left(t\right)=1,\label{T_11_0}
\end{equation}

\begin{equation}
q^{\left(1\right)}\left(t\right)=\stackrel[0]{t}{\int}f\left(\tau\right)d\tau,\label{T_11_1}
\end{equation}

\begin{equation}
q^{\left(m\right)}\left(t\right)=\stackrel[0]{t}{\int}q^{\left(m-1\right)}\left(t-\tau\right)f_{ret}\left(\tau\right)d\tau,\;m>1.\label{T_11_2}
\end{equation}

\textbf{Proof of Theorem 10. } Equality (\ref{T_11_0}) is obvious.
Let us divide the time interval $\left(0,t\right]$ into $n$
intervals $\left(0,t_{1}\right]$, $\left(t_{1},t_{2}\right]$,
$...,\left(t_{n-1},t_{n}\right]$, where $t_{0}=0$ and $t_{n}=t$.
Let $B\left(t_{i-1},t_{i}\right)=\left\{
\omega\in\Omega\vcentcolon
t_{i-1}<\tau_{B_{r}\left(a\right)}\left(\omega\right)\leq
t_{i}\right\} $ be independent events such that the trajectory of
the stochastic process $\xi\left(t,\omega\right)$ hits the domain
$B_{r}\left(a\right)$ for the first time during the time interval
$\left(t_{i-1},t_{i}\right]$. Then from the obvious relation

\[
Q_{t}^{m}\subset\stackrel[i=1]{n}{\cup}B\left(t_{i-1},t_{i}\right)
\]
we obtain

\[
Q_{t}^{m}=Q_{t}^{m}\cap
Q_{t}^{m}=Q_{t}^{m}\cap\left(\stackrel[i=1]{n}{\cup}B\left(t_{i-1},t_{i}\right)\right)=\stackrel[i=1]{n}{\cup}\left(Q_{t}^{m}\cap
B\left(t_{i-1},t_{i}\right),\right);
\]
hence the probabilities satisfy the equation

\[
\mathrm{P}\left\{ Q_{t}^{m}\right\} =\mathrm{P}\left\{ \stackrel[i=1]{n}{\cup}\left(Q_{t}^{m}\cap B\left(t_{i-1},t_{i}\right)\right)\right\} =\stackrel[i=1]{n}{\sum}\mathrm{P}\left\{ Q_{t}^{m}\cap B\left(t_{i-1},t_{i}\right)\right\}
\]

\[
=\stackrel[i=1]{n}{\sum}\mathrm{P}\left\{ Q_{t}^{m}\mid B\left(t_{i-1},t_{i}\right)\right\} \mathrm{P}\left\{ B\left(t_{i-1},t_{i}\right)\right\} .
\]
Next, the following relations hold:

\begin{equation}
\mathrm{P}\left\{ Q_{t}^{m}\right\} =\stackrel[i=1]{n}{\sum}\mathrm{P}\left\{ Q_{t-t_{i}}^{m}\mid B\left(0,t_{i}-t_{i-1}\right)\right\} \mathrm{P}\left\{ B\left(t_{i-1},t_{i}\right)\right\} ,\label{P_Q}
\end{equation}

\begin{equation}
\mathrm{P}\left\{ B\left(t_{i-1},t_{i}\right)\right\} =\left\{ \begin{array}{c}
f\left(t_{i}\right)\left(t_{i}-t_{i-1}\right)+o\left(t_{i}-t_{i-1}\right),\;m=1,\\
f_{ret}\left(t_{i}\right)\left(t_{i}-t_{i-1}\right)+o\left(t_{i}-t_{i-1}\right),\;m>1,
\end{array}\right.\label{P_Q_1}
\end{equation}

\begin{equation}
\mathrm{P}\left\{ Q_{t-t_{i}}^{m}\mid B\left(0,t_{i}-t_{i-1}\right)\right\} =\mathrm{P}\left\{ Q_{t-t_{i}}^{m-1}\right\} =q^{\left(m-1\right)}\left(t-t_{i}\right)+O\left(t_{i}-t_{i-1}\right),\label{P_Q_2}
\end{equation}

\begin{equation}
\mathrm{P}\left\{ Q_{t}^{m}\right\} =q^{\left(m\right)}\left(t\right),\label{P_Q_3}
\end{equation}
which are the consequence of the homogeneity and Markov property
of the stochastic process $\xi\left(t,\omega\right)$. Substituting
(\ref{P_Q_1})--(\ref{P_Q_3}) into (\ref{P_Q}), we obtain

\[
q^{\left(1\right)}\left(t\right)=\sum_{i=1}^{n}\left(f\left(t_{i}\right)\left(t_{i}-t_{i-1}\right)+o\left(t_{i}-t_{i-1}\right)\right),
\]

\[
q^{\left(m\right)}\left(t\right)=\sum_{i=1}^{n}\left(q^{\left(m-1\right)}\left(t-t_{i}\right)+O\left(t_{i}-t_{i-1}\right)\right)\left(f_{ret}\left(t_{i}\right)\left(t_{i}-t_{i-1}\right)+o\left(t_{i}-t_{i-1}\right)\right),
\]
and, as $n\rightarrow\infty$ and $\underset{i}{\max}\left\{
\left|t_{i}-t_{i-1}\right|\right\} \rightarrow0$, we have
(\ref{T_11_1}) and (\ref{T_11_2}), which proves Theorem 10.

The solution of equations (\ref{T_11_1}) and (\ref{T_11_2}) in
terms of the Laplace transforms has the form

\begin{equation}
Q^{\left(m\right)}\left(s\right)=\dfrac{1}{s}F\left(s\right)\left(F_{ret}\left(s\right)\right)^{m-1},\:m\geq1,\label{Q_m}
\end{equation}
where

\begin{equation}
F_{ret}\left(s\right)=1-\dfrac{1}{p^{r}\left(B_{\alpha}\left(r\right)+s\right)J_{r}\left(s\right)}\label{ret_F}
\end{equation}
is the Laplace transform of the probability density of the first
return time.

Using the properties of the function $F\left(s\right)$, as well as
the properties of the function $F_{ret}\left(s\right)$,

\[
\stackrel[0]{\infty}{\int}f_{ret}\left(t\right)dt=\underset{s\rightarrow0}{\lim}F_{ret}\left(s\right)=\left\{ \begin{array}{c}
1,\;\alpha\geq1,\\
\dfrac{p}{p^{\alpha}}\left(\dfrac{p^{\alpha}-1}{p-1}\right)^{2},\;0<\alpha<1,
\end{array}\right.
\]

\[
\stackrel[0]{\infty}{\int}tf_{ret}\left(t\right)dt=\underset{s\rightarrow0}{-\lim}\dfrac{d}{ds}F_{ret}\left(s\right)=+\infty,\;\alpha\geq1,
\]

\[
\underset{t\rightarrow0}{\lim}f_{ret}\left(t\right)=\underset{s\rightarrow\infty}{\lim}sF_{ret}\left(s\right)=0,
\]

\[
\underset{t\rightarrow\infty}{\lim}f_{ret}\left(t\right)=\underset{s\rightarrow0}{\lim}sF_{ret}\left(s\right)=0,
\]
established in \cite{ABZ_2009}, we can establish a number of
properties of the probability $q^{\left(m\right)}\left(t\right)$.
The probability $q^{\left(m\right)}\left(t\right)$ of the $m$th
hitting the domain $B_{r}\left(a\right)$ by the trajectory of the
stochastic process $\xi\left(t,\omega\right)$ within the time
interval $\left(0,t\right]$ is a nondecreasing function and
$q^{\left(m\right)}\left(t\right)\in\mathrm{C^{\infty}\left(\mathbb{R}_{+}\right)}$,
which can be easily shown from the recurrence equations
(\ref{T_11_0})--(\ref{T_11_2}) by induction. Moreover,

\[
\underset{t\rightarrow0}{\lim}q^{\left(m\right)}\left(t\right)=0,
\]

\[
\underset{t\rightarrow\infty}{\lim}q^{\left(m\right)}\left(t\right)=\left\{ \begin{array}{c}
1,\;\alpha\geq1,\\
C\left(C_{ret}\right)^{m-1},\;0<\alpha<1.
\end{array}\right.
\]
The function $\dfrac{d}{dt}q^{\left(m\right)}\left(t\right)$ --
the probability density of the $m$th hitting the domain
$B_{r}\left(a\right)$ by the trajectory of the stochastic process
$\xi\left(t,\omega\right)$ has the following properties:

\[
\underset{t\rightarrow0}{\lim}\dfrac{d}{dt}q^{\left(m\right)}\left(t\right)=0,
\]

\[
\underset{t\rightarrow\infty}{\lim}\dfrac{d}{dt}q^{\left(m\right)}\left(t\right)=0.
\]
Obviously, since the probability density function
$\dfrac{d}{dt}q^{\left(m\right)}\left(t\right)$ is continuous and
positive, it has at least one maximum.

For $\alpha\geq1$ it makes sense to speak of the average time of
the $m$th hitting the domain $B_{r}\left(a\right)$; nevertheless,
this quantity is divergent:

\[
\lim_{T\rightarrow\infty}\stackrel[0]{T}{\int}t\dfrac{d}{dt}q^{\left(m\right)}\left(t\right)dt=-\underset{s\rightarrow0}{\lim}\dfrac{d}{ds}\left(F\left(s\right)\left(F_{ret}\left(s\right)\right)\right)=+\infty.
\]
In other words, the event of the $m$th hitting is certain, but
must be waited infinitely long.

It is also easily seen that
$q^{\left(m\right)}\left(t\right)<q^{\left(m-1\right)}\left(t\right)$
for any $m=1,2,3,...$; i.e., the sequence
$q^{\left(m\right)}\left(t\right)$ monotonically decreases for any
$t.$ This follows from the fact that $Q_{t}^{m}\subset
Q_{t}^{m-1}$. Thus, for any instant of time $t$, a smaller number
of hittings the domain $B_{r}\left(a\right)$ is more probable than
a larger number.

Let us pass to the problem of finding the probability of exactly
$m$ hittings the domain $B_{r}\left(a\right)$ by the trajectories
of the stochastic process $\xi\left(t,\omega\right)$. Let
$H_{t}^{m}=\left\{ \omega\in\Omega\vcentcolon
N_{B_{r}\left(a\right)}\left(t,\omega\right)=m\right\} $ be an
event such that the trajectory of the stochastic process
$\xi\left(t,\omega\right)$ hits the domain $B_{r}\left(a\right)$
exactly $m$ times within the time interval $\left(0,t\right]$. Let
us state the problem of calculating the probability
$\mathrm{P}\left\{ H_{t}^{m}\right\}
=h^{\left(m\right)}\left(t\right)$.

\textbf{Theorem 11. } Let $f\left(t\right)$ be the probability
density function of the first passage time to the domain
$B_{r}\left(a\right)$, and let $f_{ret}\left(t\right)$ be the
probability density function of the first return time to
$B_{r}\left(a\right)$. Then the probability
$h^{\left(m\right)}\left(t\right)$ satisfies the following
recurrence equation:

\begin{equation}
h^{\left(0\right)}\left(t\right)=1-\stackrel[0]{t}{\int}f\left(\tau\right)d\tau,\label{ret_h_1}
\end{equation}

\begin{equation}
h^{\left(m\right)}\left(t\right)=\stackrel[0]{t}{\int}h^{\left(m-1\right)}\left(t-\tau\right)f_{ret}\left(\tau\right)d\tau.\label{ret_h_2}
\end{equation}

\textbf{Proof of Theorem 11.} From $H_{t}^{m}=Q_{t}^{m}\setminus
Q_{t}^{m}$ we have

\[
h^{\left(m\right)}\left(t\right)=q^{\left(m\right)}\left(t\right)-q^{\left(m+1\right)}\left(t\right),
\]
which implies the recurrence equation (\ref{ret_h_2}). Theorem 11
is proved.

The solution of equation (\ref{ret_h_2}) in terms of Laplace
transforms is given by

\[
H^{\left(0\right)}\left(s\right)=\dfrac{1}{s}\left(1-F\left(s\right)\right),
\]

\begin{equation}
H^{\left(m\right)}\left(s\right)=\dfrac{1}{s}F\left(s\right)\left(1-F_{ret}\left(s\right)\right)\left(F_{ret}\left(s\right)\right)^{m-1},\;m\geq1.\label{ret_H_L}
\end{equation}
For $t\in\mathbb{R}_{+}$, $h^{\left(m\right)}\left(t\right)$  is a
positive function. By definition, it has the following properties:
\[
h^{\left(m\right)}\left(t\right)<h^{\left(m-1\right)}\left(t\right),\;m\geq0,
\]

\[
\underset{t\rightarrow0}{\lim}h^{\left(m\right)}\left(t\right)=\underset{s\rightarrow\infty}{\lim}F\left(s\right)\left(1-F_{ret}\left(s\right)\right)\left(F_{ret}\left(s\right)\right)^{m-1}=0,\;\alpha\geq1,
\]

\[
\underset{t\rightarrow\infty}{\lim}h^{\left(m\right)}\left(t\right)=\underset{s\rightarrow0}{\lim}F\left(s\right)\left(1-F_{ret}\left(s\right)\right)\left(F_{ret}\left(s\right)\right)^{m-1}=0,\;\alpha\geq1,
\]

\[
\underset{t\rightarrow0}{\lim}h^{\left(m\right)}\left(t\right)=\underset{s\rightarrow\infty}{\lim}F\left(s\right)\left(1-F_{ret}\left(s\right)\right)\left(F_{ret}\left(s\right)\right)^{m-1}=0,\;0<\alpha<1,
\]
\[
\underset{t\rightarrow\infty}{\lim}h^{\left(m\right)}\left(t\right)=\underset{s\rightarrow0}{\lim}F\left(s\right)\left(1-F_{ret}\left(s\right)\right)\left(F_{ret}\left(s\right)\right)^{m-1}
\]

\[
=\left(\dfrac{p^{r}}{\left|a\right|_{p}}\right)^{1-\alpha}\dfrac{p^{\alpha}-1}{p^{\alpha}}\dfrac{p}{p-1}\left(1-\left(\dfrac{p^{\alpha}-1}{p-1}\right)^{2}\dfrac{p}{p^{\alpha}}\right)\left(\left(\dfrac{p^{\alpha}-1}{p-1}\right)^{2}\dfrac{p}{p^{\alpha}}\right)^{m-1},\;0<\alpha<1,
\]

\[
\underset{t\rightarrow0}{\lim}h^{\left(0\right)}\left(t\right)=\underset{s\rightarrow\infty}{\lim}\left(1-F\left(s\right)\right)=1,\;0<\alpha,
\]
\[
\underset{t\rightarrow\infty}{\lim}h^{\left(0\right)}\left(t\right)=\underset{s\rightarrow0}{\lim}\left(1-F\left(s\right)\right)=\left\{ \begin{array}{c}
0,\;\alpha\geq1,\\
1-\left(\dfrac{p^{r}}{\left|a\right|_{p}}\right)^{1-\alpha}\dfrac{p^{\alpha}-1}{p^{\alpha}}\dfrac{p}{p-1},\;0<\alpha<1.
\end{array}\right.
\]

Let us calculate the average number of hittings the domain
$B_{r}\left(a\right)$ by the trajectory of the stochastic process
$\xi\left(t,\omega\right)$ within the time interval
$\left(0,t\right]$, which, by the definition of the mathematical
expectation of a discrete random variable, is

\begin{equation}
\mu\left(t\right)=\stackrel[n=1]{\infty}{\sum}nh^{\left(n\right)}\left(t\right).\label{mean}
\end{equation}

\textbf{Theorem 12.} The average number of hittings domain
$B_{r}\left(a\right)$ by the trajectories of the stochastic
process $\xi\left(t,\omega\right)$ within the time interval
$\left(0,t\right]$ is given by the formula

\begin{equation}
\mu\left(t\right)=p^{r}B_{\alpha}\left(r\right)\stackrel[0]{t}{\int}\varepsilon\left(\tau,\left|a\right|_{p}\right)d\tau+\varepsilon\left(t,\left|a\right|_{p}\right)\label{mean_m}
\end{equation}
where
\[
\varepsilon\left(t,\left|a\right|_{p}\right)=\dfrac{1}{\left|a\right|_{p}}\left\{ \left(1-\dfrac{1}{p}\right)\stackrel[n=0]{\infty}{\sum}p^{-n}\exp\left(-\dfrac{p^{-\alpha n}}{\left|a\right|_{p}^{\alpha}}t\right)-\exp\left(-\dfrac{p^{\alpha}}{\left|a\right|_{p}^{\alpha}}t\right)\right\}
\]
is the fundamental solution of equation (\ref{p-KF}) at point
$a\in\mathbb{Q}_{p}$.

\textbf{Proof of Theorem 12.} We express (\ref{mean}) in terms of
Laplace transforms and, using (\ref{ret_H_L}), (\ref{ret_F}), and
(\ref{F(s)_sol}), obtain the following expression for the Laplace
transform of $\mu\left(t\right)$:

\begin{equation}
M\left(s\right)=\dfrac{1}{s}p^{r}\left(B_{\alpha}\left(r\right)+s\right)E\left(s,\left|a\right|_{p}\right)\label{mean_M}
\end{equation}
where $E\left(s,\left|a\right|_{p}\right)$ is the Laplace
transform of $\varepsilon\left(t,\left|a\right|_{p}\right)$.
Passing to the Laplace transform of equation (\ref{mean_M}), we
obtain (\ref{mean_m}). Theorem 12 is proved.

\textbf{Theorem 13.} The following asymptotics hold for the
function $\mu\left(t\right):$

\begin{equation}
\mu\left(t\right)=p^{r}B_{\alpha}\left(r\right)\left(1-\dfrac{1}{p}\right)t^{\tfrac{\alpha-1}{\alpha}}\stackrel[m=-\infty]{\infty}{\sum}\exp\left(\dfrac{2\pi im}{\alpha\ln p}\ln\left|a\right|_{p}^{-\alpha}t\right)\dfrac{\Gamma\left(\dfrac{1}{\alpha}-\dfrac{2\pi im}{\alpha\ln p}\right)}{\left(\alpha-1\right)\ln p+2\pi im}\left(1+o\left(1\right)\right),\;\alpha>1\label{T_13_1}
\end{equation}
\[
\mu\left(t\right)=\left(\dfrac{p^{r}}{\left|a\right|_{p}}\right)^{2}\dfrac{\left(p-1\right)^{2}}{\left(p^{1-\alpha}-1\right)\left(p^{1+\alpha}-1\right)}
\]

\begin{equation}
-\dfrac{p^{r}B_{\alpha}\left(r\right)}{\alpha\ln p}\left(1-\dfrac{1}{p}\right)t^{\tfrac{1-\alpha}{\alpha}}\stackrel[m=-\infty]{\infty}{\sum}\exp\left(\dfrac{2\pi im}{\alpha\ln p}\ln\left|a\right|_{p}^{-\alpha}t\right)\Gamma\left(\dfrac{1-\alpha}{\alpha}-\dfrac{2\pi im}{\alpha\ln p}\right)\left(1+o\left(1\right)\right),\;0<\alpha<1\label{T_13_2}
\end{equation}
\[
\mu\left(t\right)=\dfrac{p-1}{p+1}\dfrac{1}{\left|a\right|_{p}\ln p}
\]

\begin{equation}
\times\left\{ \ln\dfrac{t}{\left|a\right|_{p}}+\dfrac{\ln p}{\pi}\mathrm{Re}\stackrel[m=1]{\infty}{\sum}\dfrac{\left(-i\right)}{m}\left\{ 1-\exp\left(\dfrac{2\pi m}{\ln p}\ln\left|a\right|_{p}^{-1}t\right)\right\} \Gamma\left(1-\dfrac{2\pi im}{\ln p}\right)\right\} \left(1+o\left(1\right)\right),\;\alpha=1\label{T_13_3}
\end{equation}

\textbf{Proof of Theorem 13.} Since
\[
\stackrel[n=0]{\infty}{\sum}p^{-n}\exp\left(-\dfrac{p^{-\alpha n}}{\left|a\right|_{p}^{\alpha}}t\right)<ct^{-\tfrac{1}{\alpha}},
\]
where $c$ is a constant, it follows that

\[
\varepsilon\left(t,\left|a\right|_{p}\right)=\dfrac{1}{\left|a\right|_{p}}\left\{ \left(1-\dfrac{1}{p}\right)\stackrel[n=0]{\infty}{\sum}p^{-n}\exp\left(-\dfrac{p^{-\alpha n}}{\left|a\right|_{p}^{\alpha}}t\right)-\exp\left(-\dfrac{p^{\alpha}}{\left|a\right|_{p}^{\alpha}}t\right)\right\}
\]
\[
=\dfrac{1}{\left|a\right|_{p}}\left(1-\dfrac{1}{p}\right)\stackrel[n=0]{\infty}{\sum}p^{-n}\exp\left(-\dfrac{p^{-\alpha n}}{\left|a\right|_{p}^{\alpha}}t\right)\left(1+o\left(1\right)\right).
\]
Then from (\ref{mean_m}) we have

\begin{equation}
\mu\left(t\right)=p^{r}B_{\alpha}\left(r\right)\dfrac{1}{\left|a\right|_{p}}\left(1-\dfrac{1}{p}\right)\stackrel[0]{t}{\int}\stackrel[n=0]{\infty}{\sum}p^{-n}\exp\left(-\dfrac{p^{-\alpha n}}{\left|a\right|_{p}^{\alpha}}\tau\right)d\tau\left(1+o\left(1\right)\right).\label{mu}
\end{equation}
By Lemma 6, the asymptotics of the series
$\stackrel[n=0]{\infty}{\sum}p^{-n}\exp\left(-\dfrac{p^{-\alpha
n}}{\left|a\right|_{p}^{\alpha}}t\right)$ has the form
\begin{equation}
\stackrel[n=0]{\infty}{\sum}p^{-n}\exp\left(-\dfrac{p^{-\alpha n}}{\left|a\right|_{p}^{\alpha}}t\right)=\dfrac{1}{\alpha\ln p}\left(\dfrac{t}{\left|a\right|_{p}^{\alpha}}\right)^{-\tfrac{1}{\alpha}}\stackrel[m=-\infty]{\infty}{\sum}\left(\dfrac{t}{\left|a\right|_{p}^{\alpha}}\right)^{\tfrac{2\pi im}{\alpha\ln p}}\Gamma\left(\dfrac{1}{\alpha}-\dfrac{2\pi im}{\alpha\ln p}\right)\left(1+o\left(1\right)\right).\label{sum_1}
\end{equation}
Then (\ref{mu}) and (\ref{sum_1}) imply (\ref{T_13_1}).

To obtain the asymptotics for $0<\alpha<1$, we integrate the
series

\[
\stackrel[0]{t}{\int}\stackrel[n=0]{\infty}{\sum}p^{-n}\exp\left(-\dfrac{p^{-\alpha n}}{\left|a\right|_{p}^{\alpha}}\tau\right)d\tau=\dfrac{\left|a\right|_{p}^{\alpha}}{1-p^{-\left(1-\alpha\right)}}-\stackrel[n=0]{\infty}{\sum}p^{-\left(1-\alpha\right)n}\exp\left(-\dfrac{p^{-\alpha n}}{\left|a\right|_{p}^{\alpha}}t\right),
\]
and substitute the result into (\ref{mu}). Then, using Lemma 6, we
obtain (\ref{T_13_2}).

To find the asymptotics for $\alpha=1$, we first find the
asymptotics for $\varepsilon\left(t,\left|a\right|_{p}\right)$:
\[
\varepsilon\left(t,\left|a\right|_{p}\right)=\dfrac{1}{\left|a\right|_{p}}\left(1-\dfrac{1}{p}\right)\stackrel[n=0]{\infty}{\sum}p^{-n}\exp\left(-\dfrac{p^{-\alpha n}}{\left|a\right|_{p}^{\alpha}}t\right)\dfrac{1}{\left|a\right|_{p}}
\]

\[
=\left(1-\dfrac{1}{p}\right)\dfrac{t^{-1}}{\ln p}\stackrel[m=-\infty]{\infty}{\sum}\exp\left(\dfrac{2\pi im}{\ln p}\ln\dfrac{t}{\left|a\right|_{p}}\right)\Gamma\left(1-\dfrac{2\pi im}{\ln p}\right)\left(1+o\left(1\right)\right)
\]

\[
=\dfrac{1}{\left|a\right|_{p}}\left(1-\dfrac{1}{p}\right)\dfrac{t^{-1}}{\ln p}\left\{ 1+2\mathrm{Re}\stackrel[m=1]{\infty}{\sum}\exp\left(\dfrac{2\pi im}{\ln p}\ln\dfrac{t}{\left|a\right|_{p}}\right)\Gamma\left(1-\dfrac{2\pi im}{\ln p}\right)\right\} \left(1+o\left(1\right)\right).
\]
Substituting the last expression into (\ref{mu}) and integrating,
we obtain (\ref{T_13_3}). Theorem 13 is proved.

\section{Concluding remarks and prospects.}

The result of the present study is the general statement and
solution of the problem of finding the distribution density
function $f\left(t\right)$ for the first passage time to a ball
$B_{r}\left(a\right)$ of radius $p^{r}$ with center at an
arbitrary point $a\in\mathbb{Q}_{p}\setminus\mathbb{Z}_{p}$ by a
$p$-adic Markov stochastic process $\xi\left(t,\omega\right)$
whose distribution density function is the solution of the Cauchy
problem for the Vladimirov equation (\ref{p-KF})--(\ref{I_C}). In
this study, we have obtained, for the function $f\left(t\right)$,
Volterra integral equations of the first and second kind, as well
as an integrodifferential equation with absorbing domain. We have
proved a theorem on the equivalence of the solutions to these
equations. We have analyzed the properties of the solutions to
these equations. In addition, we have analyzed the stochastic
process $N\left(t,\omega\right)$ -- the number of hittings the
domain $B_{r}\left(a\right)$ by the trajectories of the stochastic
process $\xi\left(t,\omega\right)$ within time $\left(0,t\right]$,
obtained recurrence equations for the distribution function of
this process, and analyzed their solutions.

Note that all the results listed above have been obtained for a
stochastic process $\xi\left(t,\omega\right)$ whose probability
density function is the solution of the Cauchy problem for the
Vladimirov integrodifferential equation, which is a $p$-adic
analog of the Kolmogorov--Feller equation with kernel
$W\left(\left|x-y\right|_{p}\right)\sim\dfrac{1}{\left|x-y\right|_{p}^{\alpha+1}}$.
Nevertheless, we would like to outline a class of problems related
to a random variable such as the first passage time to a domain
$B_{r}\left(a\right)$ with the initial condition on a compact set
for other types of Markov stochastic processes. These problems may
be useful in modeling processes in the physics of polymers and
biopolymers, physics of spin glasses, and other problem of
physics, biology, and other fields. Let us list some of these
problems.

1. General homogeneous Markov stochastic process
$\xi\left(t,\omega\right)\vcentcolon\Omega\times\mathbb{R}_{+}\rightarrow\mathbb{Q}_{p}$
whose probability density function $\varphi\left(x,t\right)$ is
the solution of the Cauchy problem for a $p$-adic analog of the
Kolmogorov--Feller equation with difference kernel of general form

\begin{equation}
\dfrac{\partial}{\partial t}\varphi\left(x,t\right)=\underset{\mathbb{Q}_{p}}{\int}W\left(\left|x-y\right|_{p}\right)\left(\varphi\left(y,t\right)-\varphi\left(x,t\right)\right)d_{p}y\label{p-KF, G}
\end{equation}

\begin{equation}
\varphi\left(x,0\right)=\phi\left(x\right)\Omega\left(\left|x\right|_{p}p^{-r}\right),\label{NU,
G}
\end{equation}
where $\phi\left(x\right)$ is a bounded function on
$B_{r}\left(0\right)\equiv B_{r}$. For definiteness, consider the
problem of finding the distribution density function
$f_{ret}\left(t\right)$ of the first return time to domain
$B_{r}$. One can show, as was proved above in Theorem 3, that the
function $f_{ret}\left(t\right)$ satisfies the Volterra integral
equation of the first kind (\ref{Volt_2_ret}). In this case, the
solution of the Cauchy problem (\ref{p-KF, G}), (\ref{NU, G}) has
the form

\begin{equation}
\varphi\left(x,t\right)=\underset{\mathbb{Q}_{p}}{\int}\tilde{\phi}\left(k\right)\exp\left(-\tilde{W}\left(\left|k\right|_{p}\right)t\right)\chi\left(-kx\right)d_{p}k,\label{sol:p-KF, G}
\end{equation}
where

\[
\tilde{\phi}\left(k\right)=\underset{B_{r}}{\int}\phi\left(x\right)\chi\left(kx\right)d_{p}x,
\]

\[
\tilde{W}\left(\left|k\right|_{p}\right)=\underset{\mathbb{Q}_{p}}{\int}W\left(\left|x\right|_{p}\right)\left(1-\chi\left(kx\right)\right)d_{p}x.
\]
Here the Volterra equation of the first kind has the form

\begin{equation}
S_{B_{r}}\left(t\right)-\exp\left(-B\left(\alpha,r\right)t\right)=\stackrel[0]{t}{\int}S_{B_{r}}\left(t-\tau\right)f_{ret}\left(\tau\right)d\tau,\label{Volt_1_ret, G}
\end{equation}
where

\begin{equation}
S_{B_{r}}\left(t\right)=\underset{B_{r}}{\int}\varphi\left(x,t\right)d_{p}x=p^{r}\underset{B_{-r}}{\int}\tilde{\phi}\left(\left|k\right|_{p}\right)\exp\left(-\tilde{W}\left(\left|k\right|_{p}\right)t\right)d_{p}k\label{SP,
G}
\end{equation}
and

\[
B\left(\alpha,r\right)=p^{r}\underset{\mathbb{Q}_{p}\setminus B_{r}}{\int}W\left(\left|x\right|_{p}\right)d_{p}x.
\]
We can easily express the solution of equation (\ref{Volt_1_ret,
G}) in terms of Laplace transforms:

\[
F_{ret}\left(s\right)=1-\dfrac{1}{\left(B\left(\alpha,r\right)+s\right)J_{r}\left(s\right)},
\]
where
$S_{B_{r}}\left(t\right)=p^{r}\left(1-\dfrac{1}{p}\right)\stackrel[n=r]{\infty}{\sum}p^{-n}\exp\left(-\tilde{W}\left(p^{-n}\right)t\right)\risingdotseq
p^{r}\left(1-\dfrac{1}{p}\right)\stackrel[n=r]{\infty}{\sum}\dfrac{p^{-n}}{s+\tilde{W}\left(p^{-n}\right)}=J_{r}\left(s\right)$.
Using different kernels $W\left(\left|x-y\right|_{p}\right)$ in
equation (\ref{sol:p-KF, G}), we obtain different probability
density functions $\varphi\left(x,t\right)$ of the corresponding
stochastic processes $\xi\left(t,\omega\right)$. In
\cite{ABO_2003} we distinguished three types of kernels for
equation (\ref{p-KF, G}), which are of interest in different
applications,

\[
W\left(\left|x\right|_{p}\right)=\dfrac{\exp\left(-\alpha\left|x\right|_{p}\right)}{\left|x\right|_{p}},
\]

\[
W\left(\left|x\right|_{p}\right)=\dfrac{1}{\left|x\right|_{p}^{\alpha+1}},
\]

\[
W\left(\left|x\right|_{p}\right)=\dfrac{1}{\left|x\right|\ln^{\alpha}\left(1+\left|x\right|_{p}\right)},
\]
and analyzed large-time asymptotics of these solutions
(\ref{sol:p-KF, G}). In the context of the present study, it would
be of interest to study the problem of finding the distributions
of the first return times, first passage times, and the number of
hittings a given domain in the cases of exponential and
logarithmic kernels. Note that in \cite{Z_0,Z_1,Z_2,Z_3,Z_4} the
authors obtained a number of general results in this direction for
a random walk on the space $\mathbb{Q}_{p}^{n}$. Nevertheless, for
physical models, it would be useful to analyze in detail the
properties of the probability density function of random variables
such as the first return time
$\tau_{B_{r}}^{ret}\left(\omega\right)$ and the first passage time
$\tau_{B_{r}\left(a\right)}\left(\omega\right)$ and the
corresponding problems on the number of hittings for exponential
and logarithmic kernels even in $\mathbb{Q}_{p}.$

2. General inhomogeneous Markov stochastic process
$\xi\left(t,\omega\right)\vcentcolon\Omega\times\mathbb{R}_{+}\rightarrow\mathbb{Q}_{p}$
whose probability density function $\varphi\left(x,t\right)$ is
the solution of the Cauchy problem for a $p$-adic analog of the
Kolmogorov--Feller equation with a difference kernel of general
form and an autocatalytic source or a reaction sink:
\begin{equation}
\dfrac{\partial}{\partial t}\psi\left(x,t\right)=\underset{\mathbb{Q}_{p}}{\int}W\left(\left|x-y\right|_{p}\right)\left(\psi\left(y,t\right)-\psi\left(x,t\right)\right)d_{p}y\pm\lambda\Omega\left(\left|x-a\right|_{p}\right)\psi\left(x,t\right)\label{Gen_stock}
\end{equation}

\begin{equation}
\psi\left(x,0\right)=\varrho\left(\left|x\right|_{p}\right)\Omega\left(\left|x\right|_{p}\right).\label{Gen_stock_NC}
\end{equation}
For a function $W\left(\left|x\right|_{p}\right)$ in the power
form, such processes were analyzed in
\cite{ABKO_2002,B_2010_2,ABZ_2014,BZ_2021} in relation to the
simulation of experiments on the relaxation dynamics of protein.
For these processes it is also of interest to find the
distributions of the first return time, first passage time, and
the number of hittings a given domain for exponential, power, and
logarithmic kernels. As an illustration, we find the probability
density function $f_{ret}\left(t\right)$ of the random variable --
the first return time $\tau_{\mathbb{Z}_{p}}\left(\omega\right)$
to the domain $\mathbb{Z}_{p}$ -- for a stochastic process with
the probability density function described by the Cauchy problem
(\ref{Gen_stock}), (\ref{Gen_stock_NC}) for $a=0$ and
$\varrho\left(\left|x\right|_{p}\right)=1$. In terms of Fourier
and Laplace transforms, this Cauchy problem has the form

\[
s\tilde{\Psi}\left(k,s\right)=\Omega\left(\left|k\right|_{p}\right)-\tilde{W}\left(\left|k\right|_{p}\right)\tilde{\Psi}\left(k,s\right)\pm\lambda\underset{\mathbb{Z}_{p}}{\int}\tilde{\Psi}\left(q,s\right)d_{p}q,\;\left|k\right|_{p}\leq1,
\]

\[
\tilde{\Psi}\left(k,s\right)=0,\;\left|k\right|_{p}>1.
\]
In this case, for the function
$G\left(s\right)=\underset{\mathbb{Z}_{p}}{\int}\tilde{\Psi}\left(k,s\right)d_{p}k$
we have
\[
G\left(s\right)=-\dfrac{I\left(s\right)}{1\pm\lambda
I\left(s\right)},
\]
where

\[
I\left(s\right)=\left(1-\dfrac{1}{p}\right)\stackrel[n=0]{\infty}{\sum}\dfrac{p^{-n}}{s+\tilde{W}\left(p^{-n}\right)}.
\]
Taking into account that $G\left(s\right)$ is the Laplace
transform of the function $S_{\mathbb{Z}_{p}}\left(t\right)$, we
can show in exactly the same way as in the proof of Theorem 3 that

\[
G\left(s\right)-\dfrac{1}{B_{\alpha}+s}=G\left(s\right)F_{ret}\left(s\right),
\]
which implies

\[
F_{ret}\left(s\right)=1+\dfrac{1\pm\lambda I\left(s\right)}{\left(B_{\alpha}+s\right)I\left(s\right)}.
\]
Next, we have to find all simple poles of the function
$F_{ret}\left(s\right)$, which are determined from the solution of
the equation $\left(B_{\alpha}+s\right)I\left(s\right)=0$, and
then expand the function $F_{ret}\left(s\right)$ in partial
fractions to obtain $f_{ret}\left(t\right)$.

3. Markov stochastic processes on inhomogeneous ultrametric spaces
\cite{BZ_2015}. In contrast to homogeneous ultrametric spaces, on
inhomogeneous ultrametric spaces the number of maximal embedded
subballs for each ball may be different. The study of stochastic
processes on inhomogeneous ultrametric spaces and the related
problems of finding random variables -- the first passage (return)
time and the number of passages (returns) within a given time
interval -- offers wide possibilities for generalizing physical
models.

\section*{Data Availability Statement}

The data supporting the findings of this study are available within
the article. All other relevant source
data are available from the corresponding author upon reasonable request.

\end{document}